\documentclass[a4paper, 11pt]{article}
\usepackage[utf8]{inputenc}
\usepackage{jcappub}
\usepackage{amsfonts}
\usepackage{graphicx}
\usepackage{amssymb}
\usepackage{amsmath}
\usepackage{breqn}
\usepackage{array}
\usepackage{tikz}
\usetikzlibrary{shapes.geometric, arrows}
\tikzstyle{st} = [rectangle, rounded corners, text width = 3cm, text centered, draw = black ]
\tikzstyle{arrow} = [->,>=stealth]

\title{Vector Galileon and inflationary magnetogenesis}

\author{Debottam Nandi$^1$} \author{and S. Shankaranarayanan$^2$}
\affiliation{${}^{1}$Department of Physics, Indian Institute of
  Technology Madras, Chennai 600036, India}
\affiliation{${}^{2}$Department of Physics, Indian Institute of
  Technology Bombay, Mumbai 400076, India}

\emailAdd{debottam@physics.iitm.ac.in}
\emailAdd{shanki@iitb.ac.in}

\abstract{Cosmological inflation provides the initial conditions for the structure formation.
However, the origin of large-scale magnetic fields can not be addressed in this framework. The key issue for this long-standing problem 
is the conformal invariance of the electromagnetic (EM) field in 4-D. While many approaches have been proposed in the literature for 
breaking conformal invariance of the EM action, here, we provide a completely new way of looking at the modifications to the EM action and 
generation of primordial magnetic fields during inflation. We explicitly construct a higher derivative EM action that breaks conformal 
invariance by demanding three conditions --- theory be described by vector potential $A^{\mu}$ and its derivatives, Gauge invariance be 
satisfied, and equations of motion be linear in second derivatives of vector potential. The unique feature of our model is that appreciable 
magnetic fields are generated at small wavelengths while tiny magnetic fields are generated at large wavelengths that are consistent with current observations.}

\begin{document}

\maketitle

\section{Introduction}

Since the early days of quantum electrodynamics, higher derivative
field theories \cite{Podolsky1948,Pais1050} have been proposed to
improve the divergence structure. However, higher-derivative theories
lead to extra degrees of freedom in the system which, in consequence,
make the Hamiltonian linear in extra momentum term. Thus, the
Hamiltonian or the energy of the system is unbounded and such systems
suffer from Ostrogradsky instability \cite{Ostro,
  Woodard:2015zca}. These negative energy states can be traded by
negative norm states leading to non-unitary theories
\cite{Barth:1983hb,Simon:1990ic,Hawking:2001yt}.  The effect of extra
degrees of freedom can also be seen from the equations of motion as
the number of degrees of freedom matches the number of initial
conditions needed to solve the equations. For example, in case of
linear second order theory, the number of degrees of freedom in phase-space
is two and in configuration space, order of equation of motion is
two. In case of higher derivative theory, we obtain higher derivative
(i.e., more than 2) equation(s) of motion. Therefore, one needs extra
initial conditions to solve the differential equations which is
again can be though of as an effect of extra degrees of freedom.

Recently, it has been realized that it is possible to construct scalar
field theories whose action can have higher derivatives, however, the
equations of motion are still second order, thus automatically
avoiding the instability~\cite{Horndeski:1974wa,Nicolis2008,Deffayet2009,Deffayet:2010qz,Deffayet:2015qwa,Nandi:2015ogk,Nandi:2016pfr}. 
These models have been constructed by imposing the Galilean symmetry in the field space, 
i.e., \[\phi \rightarrow \phi + b_\mu x^\mu + c,\] where, $b, c$ are constants and
the equations of motion are invariant under the symmetry. Nicolis et al \cite{Nicolis2008} first introduced this 
model in flat space-time which later was extended to generalized curved space-time in \cite{Deffayet2009,Deffayet:2010qz},
which also coincidentally matches with Horndeski~\cite{Horndeski:1974wa}. 
These are referred to as Galilean models and do not suffer from Ostrogradsky
instabilities~\cite{Horndeski:1974wa,Nicolis2008,Deffayet2009,Deffayet:2010qz,Deffayet:2015qwa,Nandi:2015ogk,Nandi:2016pfr}. 
In order to construct Galileon models in curved space-time, one needs to add non-minimal couplings between the matter and the gravity to vector Galileon action in flat space-time. However, most recently, it has been discovered that a `simpler' Galileon model without the non-minimal coupling terms
exists in curved space-time where, equations are third order in nature
but due to an hidden second order constraint, all equations can be
reduced to second
order~\cite{Gleyzes-et-al-2015,Lin:2014jga,Gleyzes:2014qga,Gao:2014fra}.

Scalar Galilean theories have a lot in common with Lovelock theories
of gravity~\cite{Lovelock:1971yv,Lovelock:1972vz}. Lovelock theories
are obtained by imposing three conditions --- gravity must be
described by metric and its derivatives, diffeomorphism invariance and
equations of motion be quasi-linear. Using these conditions, it can be
shown that Einstein's gravity is unique in 4-D. In higher dimensions,
$R^2 - 4 R_{ab} R^{ab} + R_{abcd} R^{abcd}$ also lead to quasi-linear
equations of motions. Lovelock extensions of Einstein gravity are
shown to be free of ghost and evade problems of
Unitarity~\cite{Zumino:1985dp,Padmanabhan:2013xyr}.

Horndeski, first provided us the generalized vector-tensor theory in
curved space-time~\cite{Horndeski1981}. Recently, in a similar way,
vector Galileons with three degrees of freedom, which is often
referred to as generalized Proca theory and generalized vector
Galileons have been
constructed~\cite{Deffayet:2010zh,Heisenberg:2014rta,Hull:2015uwa,DeFelice:2016yws,Allys:2016kbq,Rodriguez:2017ckc,Heisenberg:2017mzp}. But,
the action for all these aforementioned models contains linear time
derivatives of the fields. Unfortunately, in the literature, there is
no higher derivative vector Galileon model. Unlike scalar Galileons,
there also exists a no-go theorem~\cite{Deffayet:2013tca} that states
that, higher derivative vector Galileons cannot be constructed in flat
space-time. This possess the following question: \emph{Can we construct a higher
  derivative Electromagnetic (EM) field action by demanding following
  three conditions: theory be described by vector potential $A^{\mu}$
  and its derivatives, U(1) Gauge invariance is satisfied, i.e.,
  $A_\mu \rightarrow A_{\mu} + \partial_\mu \pi$ and equations of
  motion be second order?}

We also show that the solution to the above question may also provide 
a solution to the problem of generation of the primordial magnetic field 
during inflation. Observations
indicate that magnetic fields in galaxies which are coherent on scales
of several kpc have strengths of order
$10^{-6}$~\cite{Kronberg1994,Beck:2008ty,Enqvist:1998fw}.  Recent
FERMI measurement of gamma-rays emitted by blazars seem to provide
lower bound of the order of $10^{-16}~\rm{G}$ in
voids~\cite{Neronov73,Tevecchio2010}.
	
Several mechanisms have been proposed to explain the origin of these
magnetic fields. These can be broadly classified into two: {\sl
  top-down} and {\sl bottom-up}
scenarios~\cite{Grasso:2000wj,Widrow:2002ud,Subramanian:2009fu,Subramanian:2015lua,Widrow:2011hs}. In
the bottom-up scenario, magnetic fields are first produced in stars
and propagate outwards to galaxies and intergalactic space. In the
top-down scenario, primordial magnetic field is generated in the
early-universe and the accretion of matter within stars and galaxies
amplifies the primordial magnetic field~\cite{Grasso:2000wj}. Both the
scenarios have problems. For instance, the top-down scenario can not
generate required magnetic field strength, while the bottom-up can not
generate fields with the required coherence length. The lower bound of
the magnetic fields in the voids favors top-down scenario.
	
During inflation, the conductivity becomes negligible thus allowing a
generation of large magnetic fields during this phase~\cite{1990-Kolb.Turner-Book,Bassett:2005xm}. 
However, the problem is that the standard electromagnetic (EM) action in 4-D
space-time:
	\begin{equation}
	\label{eq:SEM}
	\!\! \mathcal{S}_{SEM} = - \int \frac{d^4x}{4} \sqrt{-g} \,F_{\mu \nu} F^{\mu \nu}; F_{\mu \nu} = \partial_{\mu}A_\nu - \partial_{\nu}A_\mu
	\end{equation}
is conformally invariant and the equations of motion of the magnetic
field in FRW space-time are time
independent~\cite{Grasso:2000wj}: $$\left(\partial^2_\eta - \nabla^2
\right) (a^2 {\bf B}) = 0.$$ Thus, to generate sufficient magnetic
fields during inflation, it is necessary to break conformal invariance
of the EM action. Starting from Turner and Widrow
\cite{1990-Kolb.Turner-Book,Dolgov:1993vg,Gasperini:1995dh,Davidson:1996rw,Calzetta:1997ku,Prokopec:2001nc,Giovannini:2000dj,Durrer:2010mq,Byrnes:2011aa,Atmjeet:2013yta,Atmjeet:2014cxa,Basak:2014qea},
several authors have suggested many ways to break the conformal
invariance of the electromagnetic field by introducing (i) coupling of
the electromagnetic field with the Ricci/Riemann tensors, (ii)
non-minimal coupling of the electromagnetic field with
scalar/axion/fermionic field and (iii) compactifaction from higher
dimensional space-time.

As mentioned earlier, in this work, we show that obtaining a higher-derivative 
Electromagnetic field action that preserves gauge-invariance and equations of 
motion is second order leads to a viable candidate for primordial magnetogenesis. Thus,
our approach provides a new way of looking at the modifications to the
EM action and the primordial magnetic field generation during
inflation. 

The work is divided into two parts. In the first part, we explicitly construct a higher derivative electromagnetic
action. In doing so, we have used non-minimal covariantization
method. In addition to that, we also rely upon to a specific yet
simple line-element, i.e., FRW metric with arbitrary Lapse function to
simplify equations as for generalized curved background, it is very
difficult to evaluate such equations. We also obtain our desire first
higher derivative vector Galileon in FRW background. We explicitly
show that in flat background, it identically vanishes and being consistent 
with the no-go theorem~\cite{Deffayet:2013tca}. 
	
In the second part of our work, we implement our newly constructed
vector Galileon model in power-law and slow-roll inflationary scenarios. 
We show that, for slow-roll inflation, $E$ and $B^\prime B$-part of
the spectrum of the energy density vanishes quickly and the $B$-part
of the spectrum of the energy density remains significant. To obtain
positive energy during this time, we fix the sign of the arbitrary
constant. The immediate consequence of the fixing the sign of the
constant is that: kinetic part in the action becomes negative. The
spectrum has large blue tilt with a special feature: denominator
contains $(1-\epsilon)$ term which diverges at the end of the
inflation, providing necessary seed magnetic field. In this work, we 
use $(-,+,+,+)$ metric signature and natural units $\hbar = c = 1/(4\pi\epsilon_0) = 1$.


\section{Part I: New vector Galileon - the model}

In this section, we briefly discuss procedure to construct
scalar Galileons and then use the method to construct vector
Galileon.

\subsection{Brief discussion about constructing scalar Galileons}

In case of single scalar field theory in flat space-time, if the
Lagrangian contains second order time derivative, equation of motion
contains, in general, fourth order time derivative. However, it can
easily be shown that using Levi-Civita tensor/completely
anti-symmetric tensor, higher derivative terms are
suppressed. Consider the example:

\begin{equation}\label{eq: Scalar Galileon}
	\mathcal{L} \sim f(\partial \phi, \varphi)\,\epsilon^{\mu \nu} \,\epsilon^{\alpha \beta} \partial_{\mu \alpha}{\phi}\, \partial_{\nu \beta}{\phi},
\end{equation}
where $\epsilon^{\mu \nu}$ is an anti-symmetric tensor. As one can
see, in flat space-time, higher derivative terms in the equation of
motion vanishes as the contraction between symmetric and
anti-symmetric tensor vanishes. Moreover, the action preserves local
Lorentz invariance. In fact, the above Lagrangian matches with the
third kind of Lagrangian in ref. \cite{Nicolis2008} if the
anti-symmetric tensor part $\epsilon^{\mu \nu} \epsilon^{\alpha
  \beta}$ is replaced in terms of the flat metric, $\eta^{\mu \nu}$ as
$\eta^{\mu \alpha}\eta^{\nu \beta} - \eta^{\mu \beta} \eta^{\nu
  \alpha}$. However, in curved space-time, to maintain Lorentz
invariance, ordinary partial derivative $\partial$ is replaced by
covariant partial derivative $\nabla$. In that case, because of
connection terms in covariant derivative, higher derivative terms
appear in the equations of motion. In order to compensate those terms,
non-minimal couplings with the curvature term are added and by
appropriately fixing the coefficients of those added terms, higher
derivative terms in the equation of motion successfully omitted and
Galilean symmetry preserved~\cite{Deffayet2009}. This method of
adding and fixing non-minimal coupling terms is referred to as non-minimal
covariantization. In a similar way, higher order order Galileons can
be constructed in flat as well as curved space-time.

We use the similar method to construct the new higher derivative
vector Galileon in the next section.

\subsection{Constructing vector Galileon}
As previously mentioned in (\ref{eq: Scalar Galileon}), using the similar technique we consider the following additional term to the standard EM action (\ref{eq:SEM}):
\begin{eqnarray}\label{eq:VecGalAction}
\mathcal{S}_{VG} = \lambda\,\int \,d^4 x\, \sqrt{-g}\,
\epsilon^{\alpha \gamma \nu} \epsilon^{\mu \eta \beta}\,
\nabla_{\alpha \beta}A_\gamma \, \nabla_{\mu \nu}A_\eta
\end{eqnarray}
where $\nabla$ is covariant derivative, $\lambda$ --- whose dimension
is inverse mass square --- is the coupling constant that determines
the effect of the higher-derivative terms in the propagation of the EM
field and $\epsilon^{\alpha\beta\gamma}$ is any anti-symmetric
tensor. Note that there is a small change in the
action compared to (\ref{eq: Scalar Galileon}). There is no $f(A_\mu)$
function in the front of the action as our goal is to preserve U(1)
symmetry, i.e., $A_{\mu} \rightarrow A_\mu + \partial_\mu \pi$. In any
dimension $(\geq 3)$, the product of two anti-symmetric tensors can be
expressed as
\begin{equation} 
\label{eq:epsilontensor}
\epsilon^{\alpha\gamma\nu}\,\epsilon^{\mu \eta\beta} =
\left|\begin{array}{ccc} g^{\alpha\mu}& g^{\alpha\eta} &
g^{\alpha\beta}\\ g^{\gamma\mu} & g^{\gamma\eta} &
g^{\gamma\beta}\\ g^{\nu\mu} & g^{\nu\eta} & g^{\nu\beta}\\
\end{array}\right|.
\end{equation}
Thus, the action (\ref{eq:VecGalAction}) is a scalar. Before we
proceed, it is important to understand how the above action behaves in
the flat Minkowski space-time:
\begin{itemize}
	\item[1.] The contraction between the first and third indices
          of the anti-symmetric tensors and the derivative of the
          vector potential $\epsilon^{\alpha \cdots \nu} \epsilon^{\mu
            \cdots \beta} \nabla_{\alpha\beta} A_{\cdots} \nabla_{\mu
            \nu} A_{\cdots}$ ensures no higher derivative terms in the
          equations of motion.
	\item[2.] The covariant derivatives $\nabla$ are replaced by
          partial derivatives $\partial$. Hence, in flat space-time,
          contraction between first two indices of the anti-symmetric
          tensor and derivative of the vector potential
          $\epsilon^{\alpha \gamma \cdots} \partial_{\alpha
            \cdots}A_{\gamma}$ preserves the gauge-invariance.
	\item[3.] It can also be shown that in flat space-time, the
          action (\ref{eq:VecGalAction}) vanishes identically. Hence,
          even with the precise construction of the action that
          preserve U(1) symmetry and the equations are of quadratic
          order as well, the action does not contribute in flat
          space-time. This is the no-go theorem by Deffayet et al
          \cite{Deffayet:2013tca}.
\end{itemize} 

In curved space-time, however, action (\ref{eq:VecGalAction}) does not
vanish due to the covariant derivatives of the vector potential which
lead to extra terms in the action. This leads to dire consequences as
\begin{itemize}
	\item[1.] Connection terms are not gauge-invariant, i.e., U(1)
          symmetry is broken.
	\item[2.] These additional terms can also lead to
          higher-derivative terms in the equation of motion.
\end{itemize}

The fact that, by construction, the action (\ref{eq:VecGalAction}) is
gauge-invariant in flat space-time implies that we need to include
non-minimal coupling terms of the electromagnetic potential and its
derivatives with the Riemann/Ricci tensors and Ricci scalars
(non-minimal covariantization) as performed in case of scalar
Galileons in curved space-time. However, {\it ab initio} we do not
know the non-minimal coupling terms and, we need to consider all
possible terms. The following modification to the electromagnetic
action becomes by adding twelve non-minimal couplings become:
\begin{eqnarray}\label{eq:FullAction}
\mathcal{S}_{VEC} = \mathcal{S}_{VG} + \lambda \sum_{i=1}^{12} \mathcal{S}_i.
\end{eqnarray}
where $S_i$'s are given by

\begin{equation}\label{eq:non-minimal}
\left.
\begin{aligned}
\mathcal{S}_1 &= D_1\,\int d^4x\, \sqrt{-g}\, {g}^{\mu \nu}
        {g}^{\alpha \beta} {g}^{\gamma \delta}\, R_{\mu \nu}
        \,{\nabla}_{\alpha}{A}_{\gamma}\, {\nabla}_{\beta}{A}_{\delta}
        \\ \mathcal{S}_2 &= D_2\,\int d^4x\, \sqrt{-g}\,{g}^{\mu
          \alpha} {g}^{\nu \beta} {g}^{\gamma \delta}\, {R}_{\mu
          \nu}\, {\nabla}_{\alpha}{A}_{\gamma}\,
           {\nabla}_{\beta}{A}_{\delta} \\ \mathcal{S}_3 &= D_3\,\int
           d^4x\, \sqrt{-g}\,{g}^{\mu \nu} {g}^{\alpha \beta}
           {g}^{\gamma \delta} \,{R}_{\mu \nu}
           \,{\nabla}_{\alpha}{A}_{\beta}
           \,{\nabla}_{\gamma}{A}_{\delta} \\ \mathcal{S}_4 &=
           D_4\,\int d^4x\, \sqrt{-g}\,{g}^{\mu \nu} {g}^{\alpha
             \delta} {g}^{\gamma \beta}\, {R}_{\mu \nu}
           \,{\nabla}_{\alpha}{A}_{\beta}\,
             {\nabla}_{\gamma}{A}_{\delta} \\ \mathcal{S}_5 &=
             D_5\,\int d^4x\, \sqrt{-g}\,{g}^{\mu \gamma} {g}^{\alpha
               \beta} {g}^{\nu \delta}\, {R}_{\mu \nu}\,
             {\nabla}_{\alpha}{A}_{\beta}\,
             {\nabla}_{\gamma}{A}_{\delta} \\ \mathcal{S}_{6} &=
             D_{6}\,\int d^4x\, \sqrt{-g}\,{g}^{\mu \alpha} {g}^{\nu
               \delta} {g}^{\gamma \beta}\, {R}_{\mu \nu}\,
             {\nabla}_{\alpha}{A}_{\beta}\,
             {\nabla}_{\gamma}{A}_{\delta} \\ \mathcal{S}_{7} &=
             D_{7}\,\int d^4x\, \sqrt{-g}\,{g}^{\mu \alpha} {g}^{\nu
               \beta} {g}^{\gamma \zeta} {g}^{\delta \eta}
             \,{R}_{\alpha \beta \gamma \delta}\,
               {\nabla}_{\mu}{A}_{\nu}\, {\nabla}_{\zeta}{A}_{\eta}
               \\ \mathcal{S}_{8} &= D_{8}\,\int d^4x\,
               \sqrt{-g}\,{g}^{\mu \alpha} {g}^{\eta \beta}
                    {g}^{\gamma \zeta} {g}^{\delta \nu} \,{R}_{\alpha
                      \beta \gamma \delta}\, {\nabla}_{\mu}{A}_{\nu}
                    \,{\nabla}_{\zeta}{A}_{\eta}~~~~~~
                    \\ \mathcal{S}_{9} &= D_{9}\,\int d^4x\,
                    \sqrt{-g}\,{g}^{\alpha \beta} {g}^{\gamma \delta}
                         {g}^{\mu \nu}\, {R}_{\alpha \beta}\,
                         {R}_{\gamma \delta}\, {A}_{\mu}
                         {A}_{\nu}\\ \mathcal{S}_{10} &= D_{10}\,\int
                         d^4x\, \sqrt{-g}\,{g}^{\alpha \beta}
                         {g}^{\gamma \mu} {g}^{\delta \nu}\,
                         {R}_{\alpha \beta}\, {R}_{\gamma \delta}\,
                         {A}_{\mu} {A}_{\nu} \\ \mathcal{S}_{11} &=
                         D_{11}\,\int d^4x\, \sqrt{-g}\,{g}^{\alpha
                           \gamma} {g}^{\beta \delta} {g}^{\mu \nu}\,
                         {R}_{\alpha \beta}\, {R}_{\gamma \delta}
                         \,{A}_{\mu} {A}_{\nu}~~~~~~
                         \\ \mathcal{S}_{12} &= D_{12}\,\int d^4x\,
                         \sqrt{-g}\,{g}^{\alpha \gamma} {g}^{\beta
                           \mu} {g}^{\delta \nu}\, {R}_{\alpha
                           \beta}\, {R}_{\gamma \delta}\, {A}_{\mu}
                              {A}_{\nu}
\end{aligned}
\right\}
\end{equation}
\noindent $D_i$'s are the twelve unknown dimensionless coefficients and
$\lambda$ is the coupling constant in action
(\ref{eq:VecGalAction}). Notice that the last four terms in the
(\ref{eq:non-minimal}) look like gauge-dependent part. The reason
behind adding the gauge-dependent terms is as follows: covariant 
derivative contains gauge-dependent par as well connection term. Hence, 
the two covariant derivatives (as can be seen in (\ref{eq:VecGalAction}) as
well as the first eight terms in the (\ref{eq:non-minimal})) have the
same structure. In order to compensate those terms, we separately add
the last four terms.

\subsection{Fixing the coefficients}
Demanding that the above action is gauge-invariant in curved
space-time and that the equations of motion do not contain higher
order terms, the coefficients $D_i$'s can be fixed uniquely. However,
it is extremely difficult to evaluate such equations in general curved
space-time. Hence, for simplicity, we consider FRW line element
$ds^2 = a(\eta)^2\, (- d\eta^2 + {\text {\bf dx}}^2)$. Also for the
time being, we drop all the spatial derivatives and only concentrate
on time derivatives.

Using the FRW line element in conformal time and after performing
integration by-parts, action (\ref{eq:FullAction}) becomes,

\begin{eqnarray}\label{eq:VecGalFRWTime}
\mathcal{S}^{FRW}_{VEC} =&& \lambda\int d^4x \Bigg( E_1\,
\frac{{a^\prime}^4}{a^6} A_0{}^2 + E_2 \, \frac{{a^\prime}^4}{a^6}
\delta^{i j} A_i A_j + E_3\,\frac{a^{\prime \prime} {a^\prime}^2}{a^5}
A_0{}^2 + E_4\,\frac{a^{\prime \prime} {a^\prime}^2}{a^5} \delta^{i j}
A_i A_j + \nonumber\\ && E_5\,\frac{{a^\prime}^2}{a^4}
{A_0{}^\prime}^2 + E_6\,\frac{{a^\prime}^2}{a^4} \delta^{i j}
A_i{}^\prime A_j{}^\prime + E_7\,\frac{a^\prime a^{\prime\prime}}{a^4}
\delta^{i j} A_i A_j{}^\prime + E_8\,\frac{a^\prime
  a^{\prime\prime}}{a^4} A_0 A_0{}^\prime + E_9 \,
\frac{{a^{\prime\prime}}^2}{a^4} A_0{}^2 \nonumber \\ && + E_{10} \,
\frac{{a^{\prime\prime}}^2}{a^4} \delta^{i j}A_i a_j +
E_{11}\,\frac{a^{\prime\prime}}{a^3} {A_0{}^\prime}^2 +
E_{12}\,\frac{a^{\prime\prime}}{a^3} \delta^{i j} A_i{}^\prime
A_j{}^\prime\Bigg),
\end{eqnarray}

\noindent where the coefficients, $E_i$'s are linear functions of
$D_i$'s and are given by the relations,

\begin{equation}\label{eq:SolveE}
\left.
\begin{aligned}
&E_1 = 6 + 15 D_2 + 12 D_5 + 15 D_6 + 2 D_8 -12 D_{11} - 9 D_{12}\\
&E_2 = 22 - 13 D_2 - 3 D_6 - 5 D_8 + 12 D_{11} + D_{12} \\
&E_3 = - 12 + 24 D_1 - 3 D_2 + 24 D_3 + 24 D_4 - 3 D_6 - 3 D_8 + 18 D_{10} + 12 D_{11} + 18 D_{12} \\
&E_4 = - 16 - 12 D_1 + 5 D_2 - 12 D_4 -  D_6 +  D_8 + 6 D_{10} - 12 D_{11} + 2 D_{12} \\
&E_5 = - 3 D_2  - 3 D_5 - 3 D_{10} \\
&E_6 = - 4 + D_{7} + D_8  + 3 D_2 \\
&E_7 = - 4 + 12 D_1 + 6 D_2 + 12 D_4 + 4 D_6 + 2 D_8 \\
&E_8 = - 12 D_1 - 6 D_2 + 24 D_3  - 12 D_4 + 6 D_5 - 6 D_6 - 6 D_8 \\
&E_9 = 6 - 36 D_9 - 18 D_{10} - 12 D_{11} - 9 D_{12}\\
&E_{10} = 36 D_{9} + 6 D_{10} + 12 D_{11} + D_{12}\\
&E_{11} = 6 D_1 + 2 D_2  + 6 D_3 + 6 D_4 + 3 D_5 + 3 D_6 \\
&E_{12} =  2 -6 D_1 - 3 D_2 - D_7 - D_8 
\end{aligned}
\right\}
\end{equation}

Let us focus on the action (\ref{eq:VecGalFRWTime}). Since we have only considered time
derivatives and dropped all spatial derivatives, it may not be easy to
identify the gauge-invariant terms. However, by looking at the action
(\ref{eq:VecGalFRWTime}), it is apparent that, except fifth and sixth
terms inside in the right hand side, all other terms
are either gauge-dependent terms or may lead to higher-derivative
terms in the equations of motion. Hence, at this stage, we can safely
avoid the gauge-dependent terms and the terms that lead to higher
derivative terms in the equations of motion by setting all $E_i$'s (except $E_5$ and $E_6$) to zero and solve equations
(\ref{eq:SolveE}). This leads to

\begin{equation}\label{eq:FixingCoeff}
\left.
\begin{aligned}
&D_2 = 2,  \quad\\
&D_4 = - D_1 - D_3, \\
&D_{6} = -2 - D_5 \\
&D_{7} = - 4 - 6 D_1 - 6 D_3 - 2 D_5,\quad \quad\\
&D_{8} = 6 D_3 + 2 D_5,\\
&D_{9} = \frac{1}{12} - \frac{D_3}{2} - \frac{D_5}{12},\\
&D_{10} = -\frac{1}{6}, \\
&D_{11} = - \frac{1}{4} + \frac{3 D_3}{2} + \frac{D_5}{4}, \\
&D_{12} = 1.
\end{aligned}
\right\}
\end{equation}

As it is apparent now, we have solved $D_i$'s using the ten equations and
out of these equations, nine are independent. Hence, out
of twelve coefficients, only nine can be fixed. Moreover, it can also
be seen that, $E_5$ automatically satisfies the solution and vanishes,
and only $E_6$ survives in the action as

\begin{equation}\label{eq:VecFRWafterFixing}
S_{VEC}^{FRW} = - \lambda \left(1 + 3 D_1 \right)\int d^4x \frac{{a^\prime}^2}{a^4} \delta^{i j} A_i{}^\prime A_j{}^\prime
\end{equation}

It is also interesting to see that, $D_3$ and $D_5$ are
arbitrary parameters, and though they do not vanish, $\mathcal{S}_3$ and
$\mathcal{S}_5$ do not contribute to the action and the action
(\ref{eq:VecFRWafterFixing}) only depends on the parameter $D_1$.

Till now, we have not considered any spatial derivative. In order to
make it relatively more generalized, now we consider the action
(\ref{eq:FullAction}) not only with special derivatives but also with
FRW line element which includes arbitrary Lapse function $N(\eta)$
i.e.,
\begin{equation}
\label{eq:FRWarbitraryN}
ds^2 = - N(\eta)^2 d\eta^2 + a(\eta)^2 d{\bf x}^2.
\end{equation}
and by imposing the two necessary and sufficient conditions:
gauge-invariance and quadratic equations of motion, we try to see
whether this can constrain the the parameters further. This can lead
to one of three consequences:
\begin{itemize}
	\item[1.] It leads to extra constraint equations which exceeds
          number of unfixed parameters. In this case, constructing
          vector Galileon in curved space-time is not possible and the
          no-go theorem \cite{Deffayet:2013tca} may be also extended to curved
          space-time.
	\item[2.] It leads to extra constraint conditions but does not
          exceed the number of unfixed parameters. In that case, we
          may again fix some of the unfixed parameters and we can
          construct vector Galileon in FRW space-time.
	\item[3.] It leads to no extra conditions, hence the solution set (\ref{eq:FixingCoeff}) remains same. 
\end{itemize} 

The action (\ref{eq:FullAction}) in FRW metric with arbitrary Lapse
function is evaluated in Appendix \ref{app:Gaugefixing}. We repeat the
same procedure by imposing the desired conditions and we find that, it
does not lead to extra conditions and the action remains
gauge-invariant as well as it leads to quadratic equations of
motion. This means that, even using more generalized scenario with
spacial derivatives and arbitrary Lapse function, the solution set for
$D_i$'s (\ref{eq:FixingCoeff}) remains same and the action contains
one arbitrary parameter, $D \equiv \lambda(1 + 3 D_1)$. Because of
this reason, although we have chosen a maximally symmetric metric, we
strongly expect the relations (\ref{eq:FixingCoeff}) holds true for
arbitrary curved background as well. This is the first key result in
our work.

\section{Part II: Magnetogenesis}

Having constructed the gauge-invariant Electromagnetic action, our next 
step is to study the phenomenological consequences of this model. The modified 
electromagnetic action is
\begin{eqnarray}\label{eq:FullEM}
\mathcal{S}_{EM} = \int d^4x \sqrt{-g} \left(-\frac{1}{4}\,F_{\mu \nu}
F^{\mu \nu} + \,A_\mu J^\mu\right) +\mathcal{S}_{VEC}~~.
\end{eqnarray} 
The first two terms correspond to the standard electromagnetic
action while the last term is given by Eq. (\ref{eq:FullAction}). 
$\mathcal{S}_{VEC}$ has one unknown coupling parameter $D$ that can be fixed from observations.

As we discussed earlier, in the flat space-time, $\mathcal{S}_{VEC}$ vanishes. Thus, 
the above action reduces to standard electromagnetic action in flat space-time. 
From the equation of motion of $A_0$ in the FRW background~(\ref{eq:FRWarbitraryN}), the scalar potential
is given by:
\begin{eqnarray}\label{Permit}
\Phi \equiv - A_0 = \frac{1}{4\pi\left(1 - 4
  \,D\,H^2\right)}\,\frac{\rho(\vec{r_0})}{r}
\end{eqnarray}
where $D \equiv \lambda \,(1 + 3\, D_1)$. Thus, the effect of action
(\ref{eq:FullEM}) is to change the permitivity to $\epsilon \equiv (1
- 4 \,D\,H^2)$ where $H$ is the Hubble constant. The electrostatic
potential still goes as inverse of the distance. Permitivity being
positive provides a condition on the value of $D$. If $D$ is negative
all values are allowed, however, if $D$ is positive, then $4 D H^2 <
1$. Thus, the modified action do not have any observable consequence
in the terrestrial experiments. However, as we will show in the rest
of this work, the above modified action has important consequence in
the early Universe.

\subsection{Breaking of conformal invariance and inflationary magnetogenesis}
Having discussed the model and the effect on the Coulomb potential,
let us now look at the effects in the early Universe. Since FRW
background is conformally flat, the background gravitational field
does not produce particles in the case of standard electromagnetic
action (\ref{eq:SEM}) \cite{Birrell:1982ix}. However, the modified
action (\ref{eq:FullAction}) explicitly breaks conformal invariance
and thus may lead to production of magnetic fields and can have
significant contribution in the early Universe. Since we are
interested in the particle production and not in the vacuum polarization, 
we henceforth only consider the new term (\ref{eq:FullAction}) and ignore the 
standard EM action. As we will show this is consistent. 

Since action (\ref{eq:FullAction}) is gauge-invariant, we choose
Coulomb gauge $(A_0 = 0)$ for rest of the calculations.  In
the FRW background (\ref{eq:FRWarbitraryN}), action
(\ref{eq:FullAction}) becomes:
\begin{eqnarray}
\label{eq:TheModelFRW}
\mathcal{S}_{VEC} =  D \, \int\,d^4x  \Big[-2 \,\frac{a^\prime{}^2}{N^3\, a}\,A_i^\prime{}^2 + 2\,  \frac{a^{\prime\prime}}{N\,a^2}\, \left(\partial_i A_j\right)^2  - 2\,\frac{a^{\prime}\,N^\prime}{N^2\,a^2}\,\left(\partial_i A_j\right)^2\Big] \, .
\end{eqnarray} 

Varying the above action with respect to $A_i$ and setting $N(\eta) =
a (\eta)$, i.e., for conformal time, it leads to the following
equations of motion:

\begin{eqnarray}\label{eq:EoMASpace}
 A_i^{\prime\prime} + 2\frac{J^\prime}{J}\,A_i^\prime  - \frac{\left(aJ\right)^\prime}{\left(aJ\right)^2} \nabla^2 A_i = 0,~~ \mbox{where}~~J \equiv \frac{\mathcal{H}}{a}.
\end{eqnarray}

Fourier decomposing the vector potential
$A_i$~\cite{Subramanian:2015lua}, we get

 \begin{eqnarray}\label{eq:fourierd}
 \hat{A}_i (\eta, {\bf x}) = \sqrt{4\pi} \int \frac{d^3 {\bf
     k}}{(2\pi)^{3/2}} \sum_{\Lambda=1}^{2} \epsilon_{\Lambda i}({\bf
   k}) \Big[\hat{b}_{\bf k}^\Lambda A_k(\eta)e^{i{\bf k.x}} +
   \hat{b}_{\bf k}^{\Lambda\dagger} A^*_k(\eta) e^{- i{\bf k.x}}\Big]
 \, ,
 \end{eqnarray}
where $\Lambda$ corresponds to two orthonormal transverse
polarizations and $\epsilon_{\Lambda i}$ are the polarization
vectors. Substituting (\ref{eq:fourierd}) in (\ref{eq:EoMASpace}), 
we get
\begin{eqnarray}\label{EoMAK}
A_k^{\prime\prime} + 2\frac{J^\prime}{J}\,A_k^\prime + k^2 \frac{\left(aJ\right)^\prime}{\left(aJ\right)^2}  A_k = 0.
\end{eqnarray}
We can evaluate the vector potential at late times by fixing the initial 
state of the electromagnetic field. 

To compare with the observations, we need to evaluate the energy
density~\cite{Durrer:2010mq}. 0-0 component of the energy momentum
tensor $T_{\mu\nu}$ in the FRW background (\ref{eq:FRWarbitraryN}) is
$$ T_{0 0} = - \frac{N^2}{a^3} \frac{\delta \mathcal{L}}{\delta N} \, .$$
The energy density in conformal coordinates is:
\begin{eqnarray}
\rho \equiv - T^{0}_{0} 
=  - 6\, D \,\frac{\mathcal{H}^2}{a^6}\, \delta^{i j} A_i^\prime A_j^\prime 
+   4\, D\, \frac{\mathcal{H}^2}{a^6}\, \delta^{i k} \delta^{j l}\, \partial_i A_j \,\partial_k A_l  + 4\,D\,\frac{\mathcal{H}}{a^6}\, \delta^{i j}\, A_i{}^\prime\, \nabla^2 A_j \nonumber
\end{eqnarray}
The first term is the energy density of the Electric field $(\rho_E)$. 
Second and the third terms are the energy densities of the
magnetic field $(\rho_B)$ and $(\rho_{B.B^\prime})$, respectively:

Using the decomposition (\ref{eq:fourierd}), the electric, magnetic
part of the perturbation spectrum per logarithmic interval can be
written as:
\begin{eqnarray}\label{eq:SpectraB}
&& \mathcal{P}_{B}(k) \equiv  \frac{\mbox{d}}{\mbox{dln}k} \langle 0|\hat{\rho}_{B^2}|0 \rangle = \frac{16\, D\, \mathcal{H}^2}{\pi} \frac{k^5}{a^6} \left|A_k\right|^2 \\
 \label{eq:SpectraE}
&&  \mathcal{P}_{E}(k) \equiv  \frac{\mbox{d}}{\mbox{dln}k} \langle0|\hat{\rho}_{E^2}|0\rangle =  -\frac{24 \,D\, \mathcal{H}^2}{\pi} \frac{k^3}{a^6} \left|A_k^\prime\right|^2 \\
  \label{eq:SpectraBB}
&& \mathcal{P}_{_{B.B^\prime}}(k) \equiv  \frac{\mbox{d}}{\mbox{dln}k} \langle0|\hat{\rho}_{_{B.B^\prime}}|0\rangle = -\frac{16\, D\, \mathcal{H}}{\pi} \frac{k^5}{a^6}  A_k^\prime A_k^*~~~~.
 \end{eqnarray}
It is important to note the following: First, in the standard
electromagnetic action, the energy density is always positive and can
be written as $\left( B_iB^i + E_i E^i \right)$. However, in our case,
it is given by $ D \left( H^2 \, B_iB^i - H^2 \, E_i E^i - H
B_i^\prime B_i \right)$ and hence, it is not positive definite for
arbitrary background, however, depending on the model, it may become
positive-definite. The result may be identified as the nature of
Galileon models \cite{Kobayashi2011}.This is the second key result of
our work. Secondly, during most part of the evolution of the Universe,
electrical conductivity is high \cite{1990-Kolb.Turner-Book}, hence,
electric fields decay and do not contribute to the energy
density. This implies that $D > 0$.

Until now the analysis has been general and can be applied at any
stage of the Universe evolution. In the rest of this work, we
calculate the energy density of the electromagnetic field during
inflation. We assume that the inflation is driven a scalar field and
that the energy density of the electromagnetic field do not contribute
to the accelerated expansion during inflation. In other words, we
treat the electromagnetic field as a test field and obtain the power
spectrum.

Let us first consider power-law inflation i. e. $a(t) = a_0 t^p;
a(\eta) = a_0 \left(-\eta\right)^{1+\beta}$, where $p > 1; \beta \leq
- 2$. Note that $\beta = -2$ corresponds to de Sitter.  Substituting
$a(\eta)$ in (\ref{EoMAK}), we have:
 \begin{eqnarray}
 \mathcal{A}_k^{\prime\prime} + \left(c_s^2 k^2 - \frac{(2+\beta)(3+\beta)}{\eta^2}\right)\mathcal{A}_k = 0
 \end{eqnarray}
where $\mathcal{A}_k \equiv J(\eta)\,A_k$ and $c_s \equiv
-\frac{1}{1+\beta} > 0$. This is the third key result of this
work. The electromagnetic perturbations do not propagate at the speed
of light. This is not unusual, as the scalar perturbations in Galileon
inflation also propagate less than the speed of light
\cite{Kobayashi2010,Unnikrishnan2014}, however, the two speeds are not
the same.

During power-law inflation, speed of sound, $c_s$ is a constant and
the solution to the above differential equation is given by:
 \begin{eqnarray}\label{eq:GenSol}
 \mathcal{A}_k = \sqrt{-\eta}\left[C_1 J_{\beta+\frac{5}{2}}(-c_sk\eta) + C_2 J_{-\beta-\frac{5}{2}}(-c_sk\eta)\right].~~
 \end{eqnarray}
Imposing the initial condition in the sub-Hubble scales ($-k\eta
\rightarrow \infty$) that the field is in vacuum state corresponds to
$\mathcal{A}_k \to \frac{1}{\sqrt{2c_s k}}e^{-ic_sk\eta}$. This leads
to:
 \begin{equation}
C_1 = \sqrt{\frac{\pi}{4}}\frac{e^{i(\beta+1)\frac{\pi}{2}}}{\mbox{cos}\left(\beta\pi\right)},~~ C_2 = \sqrt{\frac{\pi}{4}}\frac{e^{-i\beta\frac{\pi}{2}}}{\mbox{cos}\left(\beta\pi\right)}.
 \end{equation}
It is important to note that for $\beta \leq -5/2$, $J_{\beta + 5/2}$
dominates, however, $J_{- (\beta + 5/2)}$ dominates for $\beta \geq
-5/2$.

From (\ref{eq:GenSol}), we obtain the spectra of the
energy-densities (\ref{eq:SpectraB}, \ref{eq:SpectraE}) and
(\ref{eq:SpectraBB}) at the crossing of the sound horizon ($c_s k_* =
a_* H_* = \frac{1+\beta}{\eta_*}$). The magnetic part of the energy
density is (Electric and $B.B^\prime$ part of the energy density are
provided in Appendix \ref{app:otherEner}):

\begin{eqnarray}\label{PBFinal}
\mathcal{P}_B &=& \frac{16 D}{\pi\,c_s^{11 + 2\beta}} \mathcal{F}_1(\beta)\,H_*^4\,\left( \frac{k}{k_*}\right)^{10+2\beta} ~~\mbox{for}~~\beta < -\frac{5}{2},~~ \mathcal{F}_1(\beta) = \frac{|C_1|^2}{2^{2\beta+5}\left(\Gamma(\beta+ 7/2)\right)^2} \nonumber \\
 &=& \frac{16 D}{\pi\,c_s^{1-2\beta}} \mathcal{F}_2(\beta)\,H_*^4 \left(\frac{k}{k_*}\right)^{-2\beta} ~~~~\,\mbox{for}~~\beta > -\frac{5}{2},~~\mathcal{F}_2(\beta) = \frac{|C_2|^2}{2^{-2\beta-5}\left(\Gamma(-\beta- 3/2)\right)^2}
\end{eqnarray}

This is the fourth key result regarding which we would like to stress
the following points: First, for $\beta = -5$, the magnetic spectra is
scale invariant. However, for $\beta = -5$, the electric field energy
density diverges. Hence, $\beta = -5$ is ruled out as that will lead
to negative energy density (since $D > 0$). Second, for $\beta \simeq
-2$, the spectra is highly blue-titled~\cite{Kahniashvili:2010wm}. To
go about understanding the consequence of the same, the energy spectra
during the slow-roll inflation~\cite{Bassett:2005xm} is given by (see
Appendix \ref{app:slow-roll}):
\begin{eqnarray}
\label{eq:PBGenerated}
\mathcal{P}_B = \frac{8\,D}{\pi\,c_s^5} \,H_*^4 \left(\frac{k}{k_*}\right)^4 \, ;
&& \qquad c_s = 1 - \epsilon_1
\end{eqnarray}
where $\epsilon_1$ is the first slow-roll
parameter~\cite{Bassett:2005xm}. It is interesting to note that in the
beginning of the inflation, $\epsilon_1 \ll 1$ and the speed of the EM
perturbations is close to unity. However, during inflation, as
$\epsilon_1$ increases, the speed of perturbations decrease, hence,
leading to larger value of the energy spectrum and near the exit of
inflation with $\epsilon \rightarrow 1$, large magnetic fields are
produced which may be sufficient for the galatic dynamo
condition. Finally, it is important to note that the power-spectrum in
our model has the same blue-tilt as that of the vacuum polarization
power-spectrum in the standard electromagnetic action. However, the
power-spectrum evaluated here is due to particle production during
inflation and depends on $D$ and
$c_s$~\cite{Turner:1987bw,Birrell:1982ix}. To fix these values and
compare with observations, we need to evolve magnetic fields to the
current epoch.

\subsection{Post inflationary evolution}
Reheating is expected to convert the energy in inflaton field to
radiation~\cite{Bassett:2005xm} and Universe for most cosmic history
has been good conductor ($\sigma \gg 1$).  Assuming instantaneous
reheating, the equation of motion of $A_i$ for large wavelength modes
is~\cite{Widrow:2002ud}:
\begin{eqnarray}
\ddot{A_i} + \frac{ \sigma + H\left(1 - 8 D \dot{H} -  4 D H^2\right)}{1 -  4 D H^2}\,\dot{A_i} = 0
\end{eqnarray}
 where $J^i  = - g^{i j} \sigma \dot{A_j}$. At late times, using Eq.~(\ref{Permit}), we have $D H^2 \ll 1$. Hence, the above equation reduces to:
\begin{eqnarray}
\ddot{A_i} +  \sigma\,  \dot{A_i} = 0 \quad \Rightarrow \quad A_i = C_1({\bf x}) t^{-\sigma\,t} + C_2({\bf x}) \, ,
\end{eqnarray} 
which is same as standard EM action~(\ref{eq:SEM}). Thus, the vector
potential $A_i$ is constant in time implying that the electric field
vanishes and magnetic field decays as $a^{-2}$. During
Radiation-dominated era, $H \propto a^{-2}$, and the energy density
corresponding to ${\cal S}_{VEC}$ decays as $a^{-6}$. However, the
energy density of the standard EM action goes as $a^{-4}$. At late
times, only EM action (\ref{eq:SEM}) contributes.

\subsection{Constraints from observations}
To compare whether the generated magnetic field (\ref{eq:PBGenerated})
has the right magnitude needed to seed galactic fields, we need to
compare $\rho_B$ with radiation background energy density
$\rho_{\gamma} \propto T^4$. This is because, the magnetic field
generated during inflation evolve as $\rho_{B} \propto
a^{-4}$~\cite{Turner:1987bw,Widrow:2002ud,Bassett:2005xm} which is
same as $\rho_\gamma$. Hence, the dimensionless quantity $r \equiv
\rho_{B}/\rho_\gamma$ remains approximately constant and provides a
convenient method to constrain the primordial magnetic
field~\cite{Turner:1987bw}.  From Eq.~(\ref{eq:PBGenerated}), we get,
 \begin{eqnarray}
 r \sim \frac{D}{c_s} 10^{-104}\, \lambda_{\text Mpc}^{-4} {\rm eV}^2 \, .
\end{eqnarray}    
Note that $D$ has dimensions of inverse mass square. The field
strength required to seed galactic fields with an efficient galactic
dynamo translates to $r \sim
10^{-34}$~\cite{Turner:1987bw,Widrow:2002ud}.  For length scales of $1
{\text Mpc}$, this translates to $D/c_s \sim 10^{70}$. Using the fact
that permitivity has to be positive, from Eq. (\ref{Permit}), we get
$D \sim 10^{-46}$ eV$^{-2}$. Thus, near the exit of inflation, $c_s
\sim 10^{-116}$.

This is the last key result of this work and we would like the stress
the following points: First, at the early epoch of inflation
$\epsilon_1 \ll 1$, implying that, $c_s \sim 1$, Hence, the energy
density of the magnetic fields generated at the early epoch of
inflation is tiny and the magnetic fields, present at decoupling and
homogeneous on scales larger than the horizon at that time is much
less than the current limit of $B \leq 10 \, nG$
\cite{Grasso:2000wj}. Second, appreciable seed magnetic fields are
generated only close to the exit of inflation. Thus, our model
naturally generates appreciable magnetic field at Mpc scale as the
modes that leave the horizon close to the exit of inflation re-enter
early during radiation epoch and an efficient dynamo mechanism can
generate the observed magnetic field. Thus, our model generates
appreciable magnetic fields \emph{only} for smaller wavelength
modes. This is the key unique feature of our model compared other
proposed models for magnetogenesis.

\section{Conclusions and Discussions}
In this work --- by demanding that the theory be described by vector
potential $A^{\mu}$ and its derivatives, Gauge invariance be
satisfied, and equations of motion be linear in second derivatives of
vector potential --- we have constructed a higher derivative
electromagnetic action that does not have ghosts and preserve U(1)
gauge invariance. This is the first higher derivative vector Galileon
model constructed as all other models in the literature contain linear
derivatives of the vector fields. We have shown that the higher order
terms vanish in the flat space-time and hence, consistent with the
no-go theorem by Deffayet et al \cite{Deffayet:2013tca}. 

We have shown
that the model breaks conformal invariance and generate magnetic field
during inflation. In doing so, we encountered an important aspect of
our model: the energy density is not positive definite for arbitrary
background and depending upon the model, it can become positive
definite. The magnetic fields generated have two key features:
First, the modes generated propagate less than the speed of light and
the speed of propagation depends on the slow-roll parameter
(\ref{eq:PBGenerated}). Second, the model generates appreciable
magnetic field for small wavelength modes ($\sim~{\rm Mpc}$) while the
model generates tiny magnetic fields for large wavelength modes. This
is an unique feature of our model compared to other models that generate
magnetic field during inflation. The energy density of the magnetic
field is appreciable only at the end of inflation and hence, our model
does not lead to any back-reaction. Since the vector Galileon field 
does not couple directly with inflaton \cite{Demozzi:2009fu}, our model ensures sufficient number of e-folds during inflation and at the same time, significant magnetic fields are generated very close to the exit of inflation. Therefore, by using this new and unique model, the tight constraint given in \cite{Demozzi:2009fu} is avoided while generating magnetic fields.

For the inflation to exit, $\epsilon_1 = 1$. Our model can generate
appreciable magnetic field near the exit and, in principle, can
provide a dynamical mechanism for the exit of inflation. This is under
investigation.

The magnetic field spectra generated in our model is blue tilted. This
should be contrasted with other models where the spectra can be fine
tuned~\cite{Durrer:2010mq}. Recently, Kahniashvili et al
\cite{Kahniashvili:2010wm}, have done a detailed analysis to place
constraints on the primordial magnetic field from the cosmological
data including models that have blue tilt. It is interesting to
investigate how the mass dispersion $\sigma(M,z)$ behaves and its
effect on the structure formation. This is currently under
investigation.

We showed that the action is gauge invariant and does not contain any
higher order derivatives for the FRW background. The assumption is
mainly driven for simplicity and applications to cosmology. Also, when
constructing the model, we initially chose a simpler situation with no
spatial derivative and then we considered more general situation with
arbitrary Lapse function and spatial derivatives of the gauge
field. In both the situations, the solution remained the same and
therefore, since the model is locally Lorentz invariant, we highly
expect the model should be applicable for arbitrary curved background
as well.

\section{Acknowledgments} We thank S. Sethi, T. R. Seshadri and K. Subramanian for useful discussions. DN is supported by CSIR fellowship. Work is supported by Max Planck-India Partner group on Gravity and Cosmology.  Further we thank
Kasper Peeters for his useful program Cadabra
\cite{Peeters:2007wn,DBLP:journals/corr/abs-cs-0608005} and useful
algebraic calculations with it.

\appendix

\section{Action in FRW space-time with arbitrary Lapse function}
\label{app:Gaugefixing}

Evaluating the equations of motion for an arbitrary metric is hard and also non-transparent.  Hence, to calculate equations of motion and thus to fix the coefficients, we consider FRW background 
$$ds^2 = - N(\eta)^2 d\eta^2 + a^2 d{\bf x}^2$$ where $N(\eta)$ is the
Lapse function. The action becomes

{\small
	\begin{eqnarray}
	&&\mathcal{L}_{SV} = 4\, \delta^{i j} {A}_{i} {\partial}_{0
            0}{{A}_{j}}\, N{}^{(-4)} a^\prime{}^{2} a{}^{(-4)} - 2\,
          \delta^{i j} {A}_{i} {A}_{j} {\delta}_{i j} N{}^{(-4)}
          a^\prime{}^{4} a{}^{(-6)} - 4\, {\delta}^{i j}{A}_{i}
          {A}_{j} a^{\prime \prime}\, N{}^{(-4)} a^\prime{}^{2}
          a{}^{(-5)} - \nonumber \\ &&8\,\delta^{i j} {A}_{i}
          {\partial}_{0}{{A}_{j}}\, N^{\prime}\, N{}^{(-5)}
          a^\prime{}^{2} a{}^{(-4)} + 4\,\delta^{i j} {A}_{i} {A}_{j}
          N^{\prime}\, N{}^{(-5)} a^\prime{}^{3} a{}^{(-5)} - 4\,
          {\delta}^{i j} {\partial}_{0}{{A}_{i}}\, a^{\prime}\,
          {\partial}_{0 0}{{A}_{j}}\, N{}^{(-4)} a{}^{(-3)} +\nonumber
          \\ && 4\, \delta^{i j} {A}_{i} {\partial}_{0}{{A}_{j}}\,
          a^{\prime}\, a^{\prime \prime}\, N{}^{(-4)} a{}^{(-4)} + 2\,
          {\delta}^{i j} {\partial}_{0}{{A}_{i}}\,
          {\partial}_{0}{{A}_{j}}\, N{}^{(-4)} a^\prime{}^{2}
          a{}^{(-4)} + 4\, {\delta}^{i j} {\partial}_{0}{{A}_{i}}\,
          {\partial}_{0}{{A}_{j}}\, N^{\prime}\, a^{\prime}\,
          N{}^{(-5)} a{}^{(-3)} + \nonumber \\ &&8\,\delta^{i j}
          {A}_{i} N^{\prime}\, {\partial}_{j}{{A}_{0}}\, N{}^{(-5)}
          a^\prime{}^{2} a{}^{(-4)} - 4\, \delta^{i j} {A}_{i}
          a^{\prime}\, {\partial}_{j}{{A}_{0}}\, a^{\prime \prime}\,
          N{}^{(-4)} a{}^{(-4)} - 4\, {\delta}^{i j}
          {\partial}_{0}{{A}_{i}}\, {\partial}_{j}{{A}_{0}}\,
          N{}^{(-4)} a^\prime{}^{2} a{}^{(-4)} - \nonumber \\ && 8\,
          {\delta}^{i j} {\partial}_{0}{{A}_{i}}\, N^{\prime}\,
          a^{\prime}\, {\partial}_{j}{{A}_{0}}\, N{}^{(-5)} a{}^{(-3)}
          + 4\, {\delta}^{i j} N^{\prime}\, a^{\prime}\,
          {\partial}_{i}{{A}_{0}}\, {\partial}_{j}{{A}_{0}}\,
          N{}^{(-5)} a{}^{(-3)} - 2\, {\delta}^{i j} {\delta}^{k l}
          {\partial}_{0 i}{{A}_{k}}\, {\partial}_{0 j}{{A}_{l}}\,
          N{}^{(-2)} a{}^{(-4)} + \nonumber \\ &&4\, {A}_{0}
          {\delta}^{i j} N^{\prime}\, a^{\prime}\, {\partial}_{0
            i}{{A}_{j}}\, N{}^{(-5)} a{}^{(-3)} - 4\, {A}_{0}
          {\delta}^{i j} a^{\prime \prime}\, {\partial}_{0
            i}{{A}_{j}}\, N{}^{(-4)} a{}^{(-3)} + 6\, {\delta}^{i j}
          {\delta}^{k l} a^{\prime}\, {\partial}_{i}{{A}_{k}}\,
          {\partial}_{0 j}{{A}_{l}}\, N{}^{(-2)} a{}^{(-5)} +
          \nonumber \\ &&6\, {A}_{0}{}^{2} N{}^{(-8)} N^\prime{}^{2}
          a^\prime{}^{2} a{}^{(-2)}
          a^{\prime \prime}\, {A}_{0}{}^{2} N{}^{(-7)} a{}^{(-2)} +
          6\, {A}_{0}{}^{2} N{}^{(-6)} a^{\prime \prime}{}^{2}
          a{}^{(-2)} - \nonumber \\ && 6\, {\delta}^{i j} {\delta}^{k
            l} {\partial}_{i}{{A}_{k}}\, {\partial}_{j}{{A}_{l}}\,
          N{}^{(-2)} a^\prime{}^{2} a{}^{(-6)} - 4\, \delta^{i j}
          {A}_{i} {\partial}_{0 j}{{A}_{0}}\, N{}^{(-4)}
          a^\prime{}^{2} a{}^{(-4)} - 4\, {\delta}^{i j} a^{\prime}\,
          {\partial}_{i}{{A}_{0}}\, {\partial}_{0 j}{{A}_{0}}\,
          N{}^{(-4)} a{}^{(-3)} + \nonumber \\ &&2\, {\delta}^{i j}
          {\partial}_{i}{{A}_{0}}\, {\partial}_{j}{{A}_{0}}\,
          N{}^{(-4)} a^\prime{}^{2} a{}^{(-4)} + 4\, {\delta}^{i j}
          {\partial}_{0}{{A}_{i}}\, a^{\prime}\, {\partial}_{0
            j}{{A}_{0}}\, N{}^{(-4)} a{}^{(-3)} - {\delta}^{i j}
          {\delta}^{k l} {\partial}_{i k}{{A}_{0}}\, {\partial}_{j
            l}{{A}_{0}}\, N{}^{(-2)} a{}^{(-4)} - \nonumber \\ &&4\,
          {A}_{0} {\delta}^{i j} N^{\prime}\, a^{\prime}\,
          {\partial}_{i j}{{A}_{0}}\, N{}^{(-5)} a{}^{(-3)} + 6\,
          {\delta}^{i j} {\delta}^{k l} {\partial}_{i}{{A}_{k}}\,
          {\partial}_{l}{{A}_{j}}\, N{}^{(-2)} a^\prime{}^{2}
          a{}^{(-6)} + {\delta}^{i j} {\delta}^{k l} {\delta}^{m n}
          {\partial}_{i k}{{A}_{m}}\, {\partial}_{j l}{{A}_{n}}\,
          a{}^{(-6)} - \nonumber \\ &&2\, {A}^{j} {\delta}^{k l}
          {\partial}_{k l}{{A}_{j}}\, N{}^{(-2)} a^\prime{}^{2}
          a{}^{(-6)} + 2\, \delta^{i j} {A}_{i} {\delta}^{k l}
          {\partial}_{j k}{{A}_{l}}\, N{}^{(-2)} a^\prime{}^{2}
          a{}^{(-6)} + {\delta}^{i j} {\delta}^{k l} {\partial}_{0
            i}{{A}_{j}}\, {\partial}_{0 k}{{A}_{l}}\, N{}^{(-2)}
          a{}^{(-4)} + \nonumber \\ &&2\, {\delta}^{i j} {\delta}^{k
            l} {\partial}_{0 i}{{A}_{0}}\, {\partial}_{j k}{{A}_{l}}\,
          N{}^{(-2)} a{}^{(-4)} - 2\, {\delta}^{i j} {\delta}^{k l}
          N^{\prime}\, {\partial}_{i}{{A}_{0}}\, {\partial}_{j
            k}{{A}_{l}}\, N{}^{(-3)} a{}^{(-4)} - {\delta}^{i j}
          {\delta}^{k l} {\delta}^{m n} {\partial}_{i k}{{A}_{j}}\,
          {\partial}_{l m}{{A}_{n}}\, a{}^{(-6)} - \nonumber \\ &&2\,
          {\delta}^{i j} {\delta}^{k l} {\partial}_{0 i}{{A}_{j}}\,
          {\partial}_{k l}{{A}_{0}}\, N{}^{(-2)} a{}^{(-4)} + 4\,
          {A}_{0} {\delta}^{i j} a^{\prime \prime}\, {\partial}_{i
            j}{{A}_{0}}\, N{}^{(-4)} a{}^{(-3)} + 4\, {\delta}^{i j}
          a^{\prime}\, {\partial}_{i}{{A}_{0}}\, {\partial}_{0
            0}{{A}_{j}}\, N{}^{(-4)} a{}^{(-3)}
          {\delta}^{i j} {\delta}^{k l} {\partial}_{0 i}{{A}_{0}}\,
          {\partial}_{k l}{{A}_{j}}\, N{}^{(-2)} a{}^{(-4)} + 2\,
          {\delta}^{i j} {\delta}^{k l} N^{\prime}\,
          {\partial}_{i}{{A}_{0}}\, {\partial}_{k l}{{A}_{j}}\,
          N{}^{(-3)} a{}^{(-4)} - 2\, {\delta}^{i j} {\delta}^{k l}
          {\partial}_{0 0}{{A}_{i}}\, {\partial}_{j k}{{A}_{l}}\,
          N{}^{(-2)} a{}^{(-4)} +\nonumber \\ && 2\, \delta^{i
            j}{A}_{i} {\delta}^{k l} a^{\prime \prime}\, {\partial}_{j
            k}{{A}_{l}}\, N{}^{(-2)} a{}^{(-5)} + 2\, {\delta}^{i j}
          {\delta}^{k l} {\partial}_{0}{{A}_{i}}\, N^{\prime}\,
          {\partial}_{j k}{{A}_{l}}\, N{}^{(-3)} a{}^{(-4)} - 2\,
          \delta^{i j} {\delta}^{k l} {A}_{i} N^{\prime}\,
          a^{\prime}\, {\partial}_{j k}{{A}_{l}}\, N{}^{(-3)}
          a{}^{(-5)} + \nonumber \\ && 2\, {\delta}^{i j} {\delta}^{k
            l} {\delta}^{m n} {\partial}_{i j}{{A}_{k}}\,
          {\partial}_{l m}{{A}_{n}}\, a{}^{(-6)} + 2\, {\delta}^{i j}
          {\delta}^{k l} {\partial}_{0 i}{{A}_{k}}\, {\partial}_{j
            l}{{A}_{0}}\, N{}^{(-2)} a{}^{(-4)} - 6\, {\delta}^{i j}
          {\delta}^{k l} a^{\prime}\, {\partial}_{i}{{A}_{k}}\,
          {\partial}_{0 l}{{A}_{j}}\, N{}^{(-2)} a{}^{(-5)} +\nonumber
          \\ && {\delta}^{i j} {\delta}^{k l} {\partial}_{0
            i}{{A}_{k}}\, {\partial}_{0 l}{{A}_{j}}\, N{}^{(-2)}
          a{}^{(-4)} - {\delta}^{i j} {\delta}^{k l} {\delta}^{m n}
          {\partial}_{i k}{{A}_{m}}\, {\partial}_{j n}{{A}_{l}}\,
          a{}^{(-6)} + 2\, {\delta}^{i j} {\delta}^{k l} {\partial}_{0
            0}{{A}_{i}}\, {\partial}_{k l}{{A}_{j}}\, N{}^{(-2)}
          a{}^{(-4)} - \nonumber \\ &&2\,\delta^{i j} {A}_{i}
          {\delta}^{k l} a^{\prime \prime}\, {\partial}_{k
            l}{{A}_{j}}\, N{}^{(-2)} a{}^{(-5)} - 2\, {\delta}^{i j}
          {\delta}^{k l} {\partial}_{0}{{A}_{i}}\, N^{\prime}\,
          {\partial}_{k l}{{A}_{j}}\, N{}^{(-3)} a{}^{(-4)} +
          2\,\delta^{i j} {\delta}^{k l} {A}_{i} N^{\prime}\,
          a^{\prime}\, {\partial}_{k l}{{A}_{j}}\, N{}^{(-3)}
          a{}^{(-5)} + \nonumber \\ && {\delta}^{i j} {\delta}^{k l}
          {\partial}_{i j}{{A}_{0}}\, {\partial}_{k l}{{A}_{0}}\,
          N{}^{(-2)} a{}^{(-4)} - {\delta}^{i j} {\delta}^{k l}
          {\delta}^{m n} {\partial}_{i j}{{A}_{k}}\, {\partial}_{m
            n}{{A}_{l}}\, a{}^{(-6)}
	\end{eqnarray}}
{\small
	\begin{eqnarray}
	&& \mathcal{L}_1 = 6\,  a^{\prime \prime}\,  N{}^{(-6)} {\partial}_{0}{{A}_{0}}\, {}^{2} a{}^{-1} - 6\,  N^{\prime}\,  a^{\prime}\,  N{}^{(-7)} {\partial}_{0}{{A}_{0}}\, {}^{2} a{}^{-1} + 6\,  N{}^{(-6)} {\partial}_{0}{{A}_{0}}\, {}^{2} a^\prime{}^{2} a{}^{(-2)} - \nonumber \\
	&&6\,  {\delta}^{i j} {\partial}_{0}{{A}_{i}}\,  {\partial}_{0}{{A}_{j}}\,  a^{\prime \prime}\,  N{}^{(-4)} a{}^{(-3)} + 6\,  {\delta}^{i j} {\partial}_{0}{{A}_{i}}\,  {\partial}_{0}{{A}_{j}}\,  N^{\prime}\,  a^{\prime}\,  N{}^{(-5)} a{}^{(-3)} - 6\,  {\delta}^{i j} {\partial}_{0}{{A}_{i}}\,  {\partial}_{0}{{A}_{j}}\,  N{}^{(-4)} a^\prime{}^{2} a{}^{(-4)} - \nonumber \\
	&&6\,  {\delta}^{i j} {\partial}_{i}{{A}_{0}}\,  {\partial}_{j}{{A}_{0}}\,  a^{\prime \prime}\,  N{}^{(-4)} a{}^{(-3)} + 6\,  {\delta}^{i j} N^{\prime}\,  a^{\prime}\,  {\partial}_{i}{{A}_{0}}\,  {\partial}_{j}{{A}_{0}}\,  N{}^{(-5)} a{}^{(-3)} - 6\,  {\delta}^{i j} {\partial}_{i}{{A}_{0}}\,  {\partial}_{j}{{A}_{0}}\,  N{}^{(-4)} a^\prime{}^{2} a{}^{(-4)} + \nonumber \\
	&&6\,  {\delta}^{i j} {\delta}^{k l} {\partial}_{i}{{A}_{k}}\,  {\partial}_{j}{{A}_{l}}\,  a^{\prime \prime}\,  N{}^{(-2)} a{}^{(-5)} - 6\,  {\delta}^{i j} {\delta}^{k l} N^{\prime}\,  a^{\prime}\,  {\partial}_{i}{{A}_{k}}\,  {\partial}_{j}{{A}_{l}}\,  N{}^{(-3)} a{}^{(-5)} + 6\,  {\delta}^{i j} {\delta}^{k l} {\partial}_{i}{{A}_{k}}\,  {\partial}_{j}{{A}_{l}}\,  N{}^{(-2)} a^\prime{}^{2} a{}^{(-6)} - \nonumber \\
	&&12\, {A}_{0}  {\partial}_{0}{{A}_{0}}\,  N^{\prime}\,  a^{\prime \prime}\,  N{}^{(-7)} a{}^{-1} + 12\, {A}_{0}  {\partial}_{0}{{A}_{0}}\,  a^{\prime}\,  N{}^{(-8)} N^\prime{}^{2} a{}^{-1} - 12\, {A}_{0}  {\partial}_{0}{{A}_{0}}\,  N^{\prime}\,  N{}^{(-7)} a^\prime{}^{2} a{}^{(-2)} + \nonumber \\
	&&6\,   {\delta}^{j i} A_j{\partial}_{0}{{A}_{i}}\,  a^{\prime}\,  a^{\prime \prime}\,  N{}^{(-4)} a{}^{(-4)} - 6\,  {\delta}^{j i} A_j {\partial}_{0}{{A}_{i}}\,  N^{\prime}\,  N{}^{(-5)} a^\prime{}^{2} a{}^{(-4)} + 6\,  {\delta}^{j i} A_j {\partial}_{0}{{A}_{i}}\,  N{}^{(-4)} a^\prime{}^{3} a{}^{(-5)} + \nonumber \\
	&&6\,   {\delta}^{j i} A_j a^{\prime}\,  {\partial}_{i}{{A}_{0}}\,  a^{\prime \prime}\,  N{}^{(-4)} a{}^{(-4)}%
	- 6\, {A}_{j}  {\delta}^{j i} N^{\prime}\,  {\partial}_{i}{{A}_{0}}\,  N{}^{(-5)} a^\prime{}^{2} a{}^{(-4)} + 6\, {A}_{j}  {\delta}^{j i} {\partial}_{i}{{A}_{0}}\,  N{}^{(-4)} a^\prime{}^{3} a{}^{(-5)} - \nonumber \\
	&&6\, {A}_{0}  {\delta}^{i j} a^{\prime}\,  {\partial}_{i}{{A}_{j}}\,  a^{\prime \prime}\,  N{}^{(-4)} a{}^{(-4)} + 6\, {A}_{0}  {\delta}^{i j} N^{\prime}\,  {\partial}_{i}{{A}_{j}}\,  N{}^{(-5)} a^\prime{}^{2} a{}^{(-4)} - 6\, {A}_{0}  {\delta}^{i j} {\partial}_{i}{{A}_{j}}\,  N{}^{(-4)} a^\prime{}^{3} a{}^{(-5)} + \nonumber \\
	&&6\, {A}_{i}  {\delta}^{i j} {\partial}_{0}{{A}_{j}}\,  a^{\prime}\,  a^{\prime \prime}\,  N{}^{(-4)} a{}^{(-4)} - 6\, {A}_{i}  {\delta}^{i j} {\partial}_{0}{{A}_{j}}\,  N^{\prime}\,  N{}^{(-5)} a^\prime{}^{2} a{}^{(-4)} + 6\, {A}_{i}  {\delta}^{i j} {\partial}_{0}{{A}_{j}}\,  N{}^{(-4)} a^\prime{}^{3} a{}^{(-5)} + \nonumber \\
	&&6\,  a^{\prime \prime}\,  {A}_{0}{}^{2} N{}^{(-8)} N^\prime{}^{2} a{}^{-1} - 6\,  a^{\prime}\,  {A}_{0}{}^{2} N{}^{(-9)} N^\prime{}^{3} a{}^{-1} + 6\,  {A}_{0}{}^{2} N{}^{(-8)} N^\prime{}^{2} a^\prime{}^{2} a{}^{(-2)} - \nonumber \\
	&&12\, {A}_{i} {A}_{j}  {\delta}^{j i} a^{\prime \prime}\,  N{}^{(-4)} a^\prime{}^{2} a{}^{(-5)} + 12\, {A}_{i} {A}_{j}  {\delta}^{j i} N^{\prime}\,  N{}^{(-5)} a^\prime{}^{3} a{}^{(-5)} - 12\, {A}_{i} {A}_{j}  {\delta}^{j i} N{}^{(-4)} a^\prime{}^{4} a{}^{(-6)} + \nonumber \\
	&&6\, {A}_{i}  {\delta}^{i j} a^{\prime}\,  {\partial}_{j}{{A}_{0}}\,  a^{\prime \prime}\,  N{}^{(-4)} a{}^{(-4)} - 6\, {A}_{i}  {\delta}^{i j} N^{\prime}\,  {\partial}_{j}{{A}_{0}}\,  N{}^{(-5)} a^\prime{}^{2} a{}^{(-4)} + 6\, {A}_{i}  {\delta}^{i j} {\partial}_{j}{{A}_{0}}\,  N{}^{(-4)} a^\prime{}^{3} a{}^{(-5)} - \nonumber \\
	&&6\, {A}_{0}  {\delta}^{i j} a^{\prime}\,  {\partial}_{j}{{A}_{i}}\,  a^{\prime \prime}\,  N{}^{(-4)} a{}^{(-4)} + 6\, {A}_{0}  {\delta}^{i j} N^{\prime}\,  {\partial}_{j}{{A}_{i}}\,  N{}^{(-5)} a^\prime{}^{2} a{}^{(-4)} - 6\, {A}_{0}  {\delta}^{i j} {\partial}_{j}{{A}_{i}}\,  N{}^{(-4)} a^\prime{}^{3} a{}^{(-5)}%
	+ \nonumber \\
	&&18\,  a^{\prime \prime}\,  {A}_{0}{}^{2} N{}^{(-6)} a^\prime{}^{2} a{}^{(-3)} - 18\,  N^{\prime}\,  {A}_{0}{}^{2} N{}^{(-7)} a^\prime{}^{3} a{}^{(-3)} + 18\, {A}_{0}{}^{2} N{}^{(-6)} a^\prime{}^{4} a{}^{(-4)} 
	\end{eqnarray}
	\begin{eqnarray}
	&& \mathcal{L}_2 = 3\,  a^{\prime \prime}\,  N{}^{(-6)} {\partial}_{0}{{A}_{0}}\, {}^{2} a{}^{-1} - 3\,  N^{\prime}\,  a^{\prime}\,  N{}^{(-7)} {\partial}_{0}{{A}_{0}}\, {}^{2} a{}^{-1} - 3\,  {\delta}^{i j} {\partial}_{0}{{A}_{i}}\,  {\partial}_{0}{{A}_{j}}\,  a^{\prime \prime}\,  N{}^{(-4)} a{}^{(-3)} + \nonumber \\
	&&3\,  {\delta}^{i j} {\partial}_{0}{{A}_{i}}\,  {\partial}_{0}{{A}_{j}}\,  N^{\prime}\,  a^{\prime}\,  N{}^{(-5)} a{}^{(-3)} +  {\delta}^{i j} N^{\prime}\,  a^{\prime}\,  {\partial}_{i}{{A}_{0}}\,  {\partial}_{j}{{A}_{0}}\,  N{}^{(-5)} a{}^{(-3)} - 2\,  {\delta}^{i j} {\partial}_{i}{{A}_{0}}\,  {\partial}_{j}{{A}_{0}}\,  N{}^{(-4)} a^\prime{}^{2} a{}^{(-4)} - \nonumber \\
	&& {\delta}^{i j} {\partial}_{i}{{A}_{0}}\,  {\partial}_{j}{{A}_{0}}\,  a^{\prime \prime}\,  N{}^{(-4)} a{}^{(-3)} -  {\delta}^{i j} {\delta}^{k l} N^{\prime}\,  a^{\prime}\,  {\partial}_{i}{{A}_{l}}\,  {\partial}_{j}{{A}_{k}}\,  N{}^{(-3)} a{}^{(-5)} + 4\,  {\delta}^{i j} {\delta}^{k l} {\partial}_{i}{{A}_{l}}\,  {\partial}_{j}{{A}_{k}}\,  N{}^{(-2)} a^\prime{}^{2} a{}^{(-6)} +\nonumber \\
	&&  {\delta}^{i j} {\delta}^{k l} {\partial}_{i}{{A}_{l}}\,  {\partial}_{j}{{A}_{k}}\,  a^{\prime \prime}\,  N{}^{(-2)} a{}^{(-5)} - 2\,  {\delta}^{i j} {\delta}^{k l} {\partial}_{i}{{A}_{k}}\,  {\partial}_{j}{{A}_{l}}\,  N{}^{(-2)} a^\prime{}^{2} a{}^{(-6)} - 6\, {A}_{0}  {\partial}_{0}{{A}_{0}}\,  N^{\prime}\,  a^{\prime \prime}\,  N{}^{(-7)} a{}^{-1} +\nonumber \\
	&& 6\, {A}_{0}  {\partial}_{0}{{A}_{0}}\,  a^{\prime}\,  N{}^{(-8)} N^\prime{}^{2} a{}^{-1} + 3\, {A}_{j}  {\delta}^{j i} {\partial}_{0}{{A}_{i}}\,  a^{\prime}\,  a^{\prime \prime}\,  N{}^{(-4)} a{}^{(-4)} - 3\, {A}_{j}  {\delta}^{j i} {\partial}_{0}{{A}_{i}}\,  N^{\prime}\,  N{}^{(-5)} a^\prime{}^{2} a{}^{(-4)} - \nonumber \\
	&&{A}_{j}  {\delta}^{j i} N^{\prime}\,  {\partial}_{i}{{A}_{0}}\,  N{}^{(-5)} a^\prime{}^{2} a{}^{(-4)} + 2\, {A}_{j}  {\delta}^{j i} {\partial}_{i}{{A}_{0}}\,  N{}^{(-4)} a^\prime{}^{3} a{}^{(-5)} + {A}_{j}  {\delta}^{j i} a^{\prime}\,  {\partial}_{i}{{A}_{0}}\,  a^{\prime \prime}\,  N{}^{(-4)} a{}^{(-4)} + \nonumber \\
	&&{A}_{0}  {\delta}^{i j} N^{\prime}\,  {\partial}_{j}{{A}_{i}}\,  N{}^{(-5)} a^\prime{}^{2} a{}^{(-4)}%
	- 2\, {A}_{0}  {\delta}^{i j} {\partial}_{j}{{A}_{i}}\,  N{}^{(-4)} a^\prime{}^{3} a{}^{(-5)} - {A}_{0}  {\delta}^{i j} a^{\prime}\,  {\partial}_{j}{{A}_{i}}\,  a^{\prime \prime}\,  N{}^{(-4)} a{}^{(-4)} - \nonumber \\
	&&2\, {A}_{0}  {\delta}^{i j} {\partial}_{i}{{A}_{j}}\,  N{}^{(-4)} a^\prime{}^{3} a{}^{(-5)} + 3\, {A}_{i}  {\delta}^{i j} {\partial}_{0}{{A}_{j}}\,  a^{\prime}\,  a^{\prime \prime}\,  N{}^{(-4)} a{}^{(-4)} - 3\, {A}_{i}  {\delta}^{i j} {\partial}_{0}{{A}_{j}}\,  N^{\prime}\,  N{}^{(-5)} a^\prime{}^{2} a{}^{(-4)} + \nonumber \\
	&&3\,  a^{\prime \prime}\,  {A}_{0}{}^{2} N{}^{(-8)} N^\prime{}^{2} a{}^{-1} - 3\,  a^{\prime}\,  {A}_{0}{}^{2} N{}^{(-9)} N^\prime{}^{3} a{}^{-1} - 3\, {A}_{i} {A}_{j}  {\delta}^{j i} a^{\prime \prime}\,  N{}^{(-4)} a^\prime{}^{2} a{}^{(-5)} + \nonumber \\
	&&3\, {A}_{i} {A}_{j}  {\delta}^{j i} N^{\prime}\,  N{}^{(-5)} a^\prime{}^{3} a{}^{(-5)} - {A}_{i}  {\delta}^{i j} N^{\prime}\,  {\partial}_{j}{{A}_{0}}\,  N{}^{(-5)} a^\prime{}^{2} a{}^{(-4)} + 2\, {A}_{i}  {\delta}^{i j} {\partial}_{j}{{A}_{0}}\,  N{}^{(-4)} a^\prime{}^{3} a{}^{(-5)} + \nonumber \\
	&&{A}_{i}  {\delta}^{i j} a^{\prime}\,  {\partial}_{j}{{A}_{0}}\,  a^{\prime \prime}\,  N{}^{(-4)} a{}^{(-4)} + {A}_{i} {A}_{j}  {\delta}^{i j} N^{\prime}\,  N{}^{(-5)} a^\prime{}^{3} a{}^{(-5)} - 4\, {A}_{i} {A}_{j}  {\delta}^{i j} N{}^{(-4)} a^\prime{}^{4} a{}^{(-6)} - \nonumber \\
	&&{A}_{i} {A}_{j}  {\delta}^{i j} a^{\prime \prime}\,  N{}^{(-4)} a^\prime{}^{2} a{}^{(-5)} + 2\, {A}_{i} {A}_{j}  {\delta}^{j i} N{}^{(-4)} a^\prime{}^{4} a{}^{(-6)} + {A}_{0}  {\delta}^{i j} N^{\prime}\,  {\partial}_{i}{{A}_{j}}\,  N{}^{(-5)} a^\prime{}^{2} a{}^{(-4)} - \nonumber \\
	&&{A}_{0}  {\delta}^{i j} a^{\prime}\,  {\partial}_{i}{{A}_{j}}\,  a^{\prime \prime}\,  N{}^{(-4)} a{}^{(-4)} - 3\,  N^{\prime}\,  {A}_{0}{}^{2} N{}^{(-7)} a^\prime{}^{3} a{}^{(-3)} + 6\,  {A}_{0}{}^{2} N{}^{(-6)} a^\prime{}^{4} a{}^{(-4)}%
	+ \nonumber \\
	&&3\,  a^{\prime \prime}\,  {A}_{0}{}^{2} N{}^{(-6)} a^\prime{}^{2} a{}^{(-3)}
	\end{eqnarray}
}
{\small
	\begin{eqnarray}
	&& \mathcal{L}_3 = 6\,  a^{\prime \prime}\,  N{}^{(-6)} {\partial}_{0}{{A}_{0}}\, {}^{2} a{}^{-1} - 6\,  N^{\prime}\,  a^{\prime}\,  N{}^{(-7)} {\partial}_{0}{{A}_{0}}\, {}^{2} a{}^{-1} + 6\,  N{}^{(-6)} {\partial}_{0}{{A}_{0}}\, {}^{2} a^\prime{}^{2} a{}^{(-2)} - \nonumber \\
	&&12\,  {\delta}^{i j} {\partial}_{0}{{A}_{0}}\,  {\partial}_{i}{{A}_{j}}\,  a^{\prime \prime}\,  N{}^{(-4)} a{}^{(-3)} + 12\,  {\delta}^{i j} {\partial}_{0}{{A}_{0}}\,  N^{\prime}\,  a^{\prime}\,  {\partial}_{i}{{A}_{j}}\,  N{}^{(-5)} a{}^{(-3)} - 12\,  {\delta}^{i j} {\partial}_{0}{{A}_{0}}\,  {\partial}_{i}{{A}_{j}}\,  N{}^{(-4)} a^\prime{}^{2} a{}^{(-4)} + \nonumber \\
	&&6\,  {\delta}^{i j} {\delta}^{k l} {\partial}_{i}{{A}_{j}}\,  {\partial}_{k}{{A}_{l}}\,  a^{\prime \prime}\,  N{}^{(-2)} a{}^{(-5)} - 6\,  {\delta}^{i j} {\delta}^{k l} N^{\prime}\,  a^{\prime}\,  {\partial}_{i}{{A}_{j}}\,  {\partial}_{k}{{A}_{l}}\,  N{}^{(-3)} a{}^{(-5)} + 6\,  {\delta}^{i j} {\delta}^{k l} {\partial}_{i}{{A}_{j}}\,  {\partial}_{k}{{A}_{l}}\,  N{}^{(-2)} a^\prime{}^{2} a{}^{(-6)} - \nonumber \\
	&&12\, {A}_{0}  {\partial}_{0}{{A}_{0}}\,  N^{\prime}\,  a^{\prime \prime}\,  N{}^{(-7)} a{}^{-1} + 12\, {A}_{0}  {\partial}_{0}{{A}_{0}}\,  a^{\prime}\,  N{}^{(-8)} N^\prime{}^{2} a{}^{-1} - 48\, {A}_{0}  {\partial}_{0}{{A}_{0}}\,  N^{\prime}\,  N{}^{(-7)} a^\prime{}^{2} a{}^{(-2)} + \nonumber \\
	&&12\, {A}_{0}  {\delta}^{i j} N^{\prime}\,  {\partial}_{i}{{A}_{j}}\,  a^{\prime \prime}\,  N{}^{(-5)} a{}^{(-3)} - 12\, {A}_{0}  {\delta}^{i j} a^{\prime}\,  {\partial}_{i}{{A}_{j}}\,  N{}^{(-6)} N^\prime{}^{2} a{}^{(-3)} + 48\, {A}_{0}  {\delta}^{i j} N^{\prime}\,  {\partial}_{i}{{A}_{j}}\,  N{}^{(-5)} a^\prime{}^{2} a{}^{(-4)} + \nonumber \\
	&& 36\, {A}_{0}  {\partial}_{0}{{A}_{0}}\,  a^{\prime}\,  a^{\prime \prime}\,  N{}^{(-6)} a{}^{(-2)} + 36\, {A}_{0}  {\partial}_{0}{{A}_{0}}\,  N{}^{(-6)} a^\prime{}^{3} a{}^{(-3)} - 36\, {A}_{0}  {\delta}^{i j} a^{\prime}\,  {\partial}_{i}{{A}_{j}}\,  a^{\prime \prime}\,  N{}^{(-4)} a{}^{(-4)} - \nonumber \\
	&&36\, {A}_{0}  {\delta}^{i j} {\partial}_{i}{{A}_{j}}\,  N{}^{(-4)} a^\prime{}^{3} a{}^{(-5)}%
	+ 6\,  a^{\prime \prime}\,  {A}_{0}{}^{2} N{}^{(-8)} N^\prime{}^{2} a{}^{-1} - 6\,  a^{\prime}\,  {A}_{0}{}^{2} N{}^{(-9)} N^\prime{}^{3} a{}^{-1} + \nonumber \\
	&&42\,  {A}_{0}{}^{2} N{}^{(-8)} N^\prime{}^{2} a^\prime{}^{2} a{}^{(-2)} - 36\,  N^{\prime}\,  a^{\prime}\,  a^{\prime \prime}\,  {A}_{0}{}^{2} N{}^{(-7)} a{}^{(-2)} - 90\,  N^{\prime}\,  {A}_{0}{}^{2} N{}^{(-7)} a^\prime{}^{3} a{}^{(-3)} + \nonumber \\
	&&54\,  a^{\prime \prime}\,  {A}_{0}{}^{2} N{}^{(-6)} a^\prime{}^{2} a{}^{(-3)} + 54\,  {A}_{0}{}^{2} N{}^{(-6)} a^\prime{}^{4} a{}^{(-4)}
	\end{eqnarray}
}
{\small
	\begin{eqnarray}
	&&\mathcal{L}_4 = 6\,  a^{\prime \prime}\,  N{}^{(-6)} {\partial}_{0}{{A}_{0}}\, {}^{2} a{}^{-1} - 6\,  N^{\prime}\,  a^{\prime}\,  N{}^{(-7)} {\partial}_{0}{{A}_{0}}\, {}^{2} a{}^{-1} + 6\,  N{}^{(-6)} {\partial}_{0}{{A}_{0}}\, {}^{2} a^\prime{}^{2} a{}^{(-2)} - \nonumber \\
	&&12\,  {\delta}^{i j} {\partial}_{0}{{A}_{j}}\,  {\partial}_{i}{{A}_{0}}\,  a^{\prime \prime}\,  N{}^{(-4)} a{}^{(-3)} + 12\,  {\delta}^{i j} {\partial}_{0}{{A}_{j}}\,  N^{\prime}\,  a^{\prime}\,  {\partial}_{i}{{A}_{0}}\,  N{}^{(-5)} a{}^{(-3)} - 12\,  {\delta}^{i j} {\partial}_{0}{{A}_{j}}\,  {\partial}_{i}{{A}_{0}}\,  N{}^{(-4)} a^\prime{}^{2} a{}^{(-4)} + \nonumber \\
	&&6\,  {\delta}^{i j} {\delta}^{k l} {\partial}_{i}{{A}_{l}}\,  {\partial}_{k}{{A}_{j}}\,  a^{\prime \prime}\,  N{}^{(-2)} a{}^{(-5)} - 6\,  {\delta}^{i j} {\delta}^{k l} N^{\prime}\,  a^{\prime}\,  {\partial}_{i}{{A}_{l}}\,  {\partial}_{k}{{A}_{j}}\,  N{}^{(-3)} a{}^{(-5)} + 6\,  {\delta}^{i j} {\delta}^{k l} {\partial}_{i}{{A}_{l}}\,  {\partial}_{k}{{A}_{j}}\,  N{}^{(-2)} a^\prime{}^{2} a{}^{(-6)} - \nonumber \\
	&&12\, {A}_{0}  {\partial}_{0}{{A}_{0}}\,  N^{\prime}\,  a^{\prime \prime}\,  N{}^{(-7)} a{}^{-1} + 12\, {A}_{0}  {\partial}_{0}{{A}_{0}}\,  a^{\prime}\,  N{}^{(-8)} N^\prime{}^{2} a{}^{-1} - 12\, {A}_{0}  {\partial}_{0}{{A}_{0}}\,  N^{\prime}\,  N{}^{(-7)} a^\prime{}^{2} a{}^{(-2)} + \nonumber \\
	&&6\, {A}_{j}  {\delta}^{j i} a^{\prime}\,  {\partial}_{i}{{A}_{0}}\,  a^{\prime \prime}\,  N{}^{(-4)} a{}^{(-4)} - 6\, {A}_{j}  {\delta}^{j i} N^{\prime}\,  {\partial}_{i}{{A}_{0}}\,  N{}^{(-5)} a^\prime{}^{2} a{}^{(-4)} + 6\, {A}_{j}  {\delta}^{j i} {\partial}_{i}{{A}_{0}}\,  N{}^{(-4)} a^\prime{}^{3} a{}^{(-5)} + \nonumber \\
	&&6\, {A}_{j}  {\delta}^{j i} {\partial}_{0}{{A}_{i}}\,  a^{\prime}\,  a^{\prime \prime}\,  N{}^{(-4)} a{}^{(-4)} - 6\, {A}_{j}  {\delta}^{j i} {\partial}_{0}{{A}_{i}}\,  N^{\prime}\,  N{}^{(-5)} a^\prime{}^{2} a{}^{(-4)} + 6\, {A}_{j}  {\delta}^{j i} {\partial}_{0}{{A}_{i}}\,  N{}^{(-4)} a^\prime{}^{3} a{}^{(-5)} - \nonumber \\
	&&12\, {A}_{0}  {\delta}^{i j} a^{\prime}\,  {\partial}_{i}{{A}_{j}}\,  a^{\prime \prime}\,  N{}^{(-4)} a{}^{(-4)}%
	+ 12\, {A}_{0}  {\delta}^{i j} N^{\prime}\,  {\partial}_{i}{{A}_{j}}\,  N{}^{(-5)} a^\prime{}^{2} a{}^{(-4)} - 12\, {A}_{0}  {\delta}^{i j} {\partial}_{i}{{A}_{j}}\,  N{}^{(-4)} a^\prime{}^{3} a{}^{(-5)} + \nonumber \\
	&&6\, {A}_{i}  {\delta}^{i j} a^{\prime}\,  {\partial}_{j}{{A}_{0}}\,  a^{\prime \prime}\,  N{}^{(-4)} a{}^{(-4)} - 6\, {A}_{i}  {\delta}^{i j} N^{\prime}\,  {\partial}_{j}{{A}_{0}}\,  N{}^{(-5)} a^\prime{}^{2} a{}^{(-4)} + 6\, {A}_{i}  {\delta}^{i j} {\partial}_{j}{{A}_{0}}\,  N{}^{(-4)} a^\prime{}^{3} a{}^{(-5)} + \nonumber \\
	&&6\,  a^{\prime \prime}\,  {A}_{0}{}^{2} N{}^{(-8)} N^\prime{}^{2} a{}^{-1} - 6\,  a^{\prime}\,  {A}_{0}{}^{2} N{}^{(-9)} N^\prime{}^{3} a{}^{-1} + 6\,  {A}_{0}{}^{2} N{}^{(-8)} N^\prime{}^{2} a^\prime{}^{2} a{}^{(-2)} - \nonumber \\
	&&6\, {A}_{i} {A}_{j}  {\delta}^{i j} a^{\prime \prime}\,  N{}^{(-4)} a^\prime{}^{2} a{}^{(-5)} + 6\, {A}_{i} {A}_{j}  {\delta}^{i j} N^{\prime}\,  N{}^{(-5)} a^\prime{}^{3} a{}^{(-5)} - 6\, {A}_{i} {A}_{j}  {\delta}^{i j} N{}^{(-4)} a^\prime{}^{4} a{}^{(-6)} + \nonumber \\
	&&6\, {A}_{i}  {\delta}^{i j} {\partial}_{0}{{A}_{j}}\,  a^{\prime}\,  a^{\prime \prime}\,  N{}^{(-4)} a{}^{(-4)} - 6\, {A}_{i}  {\delta}^{i j} {\partial}_{0}{{A}_{j}}\,  N^{\prime}\,  N{}^{(-5)} a^\prime{}^{2} a{}^{(-4)} + 6\, {A}_{i}  {\delta}^{i j} {\partial}_{0}{{A}_{j}}\,  N{}^{(-4)} a^\prime{}^{3} a{}^{(-5)} - \nonumber \\
	&&6\, {A}_{i} {A}_{j}  {\delta}^{j i} a^{\prime \prime}\,  N{}^{(-4)} a^\prime{}^{2} a{}^{(-5)} + 6\, {A}_{i} {A}_{j}  {\delta}^{j i} N^{\prime}\,  N{}^{(-5)} a^\prime{}^{3} a{}^{(-5)} - 6\, {A}_{i} {A}_{j}  {\delta}^{j i} N{}^{(-4)} a^\prime{}^{4} a{}^{(-6)} + \nonumber \\
	&&18\,  a^{\prime \prime}\,  {A}_{0}{}^{2} N{}^{(-6)} a^\prime{}^{2} a{}^{(-3)} - 18\,  N^{\prime}\,  {A}_{0}{}^{2} N{}^{(-7)} a^\prime{}^{3} a{}^{(-3)} + 18\,  {A}_{0}{}^{2} N{}^{(-6)} a^\prime{}^{4} a{}^{(-4)}
	\end{eqnarray}}
{\small
	\begin{eqnarray}
	&& \mathcal{L}_5 = 3\,  a^{\prime \prime}\,  N{}^{(-6)} {\partial}_{0}{{A}_{0}}\, {}^{2} a{}^{-1} - 3\,  N^{\prime}\,  a^{\prime}\,  N{}^{(-7)} {\partial}_{0}{{A}_{0}}\, {}^{2} a{}^{-1} +  {\delta}^{i j} {\partial}_{0}{{A}_{0}}\,  N^{\prime}\,  a^{\prime}\,  {\partial}_{j}{{A}_{i}}\,  N{}^{(-5)} a{}^{(-3)} - \nonumber \\
	&&4\,  {\delta}^{i j} {\partial}_{0}{{A}_{0}}\,  {\partial}_{j}{{A}_{i}}\,  N{}^{(-4)} a^\prime{}^{2} a{}^{(-4)} -  {\delta}^{i j} {\partial}_{0}{{A}_{0}}\,  {\partial}_{j}{{A}_{i}}\,  a^{\prime \prime}\,  N{}^{(-4)} a{}^{(-3)} + 2\,  {\delta}^{i j} {\partial}_{0}{{A}_{0}}\,  {\partial}_{i}{{A}_{j}}\,  N{}^{(-4)} a^\prime{}^{2} a{}^{(-4)} - \nonumber \\
	&&3\,  {\delta}^{i j} {\partial}_{0}{{A}_{0}}\,  {\partial}_{i}{{A}_{j}}\,  a^{\prime \prime}\,  N{}^{(-4)} a{}^{(-3)} + 3\,  {\delta}^{i j} {\partial}_{0}{{A}_{0}}\,  N^{\prime}\,  a^{\prime}\,  {\partial}_{i}{{A}_{j}}\,  N{}^{(-5)} a{}^{(-3)} -  {\delta}^{i j} {\delta}^{k l} N^{\prime}\,  a^{\prime}\,  {\partial}_{j}{{A}_{i}}\,  {\partial}_{k}{{A}_{l}}\,  N{}^{(-3)} a{}^{(-5)} +\nonumber \\
	&& 4\,  {\delta}^{i j} {\delta}^{k l} {\partial}_{j}{{A}_{i}}\,  {\partial}_{k}{{A}_{l}}\,  N{}^{(-2)} a^\prime{}^{2} a{}^{(-6)} +  {\delta}^{i j} {\delta}^{k l} {\partial}_{j}{{A}_{i}}\,  {\partial}_{k}{{A}_{l}}\,  a^{\prime \prime}\,  N{}^{(-2)} a{}^{(-5)} - 2\,  {\delta}^{i j} {\delta}^{k l} {\partial}_{i}{{A}_{j}}\,  {\partial}_{k}{{A}_{l}}\,  N{}^{(-2)} a^\prime{}^{2} a{}^{(-6)} - \nonumber \\
	&&6\, {A}_{0}  {\partial}_{0}{{A}_{0}}\,  N^{\prime}\,  a^{\prime \prime}\,  N{}^{(-7)} a{}^{-1} + 6\, {A}_{0}  {\partial}_{0}{{A}_{0}}\,  a^{\prime}\,  N{}^{(-8)} N^\prime{}^{2} a{}^{-1} + 3\, {A}_{0}  {\delta}^{i j} N^{\prime}\,  {\partial}_{i}{{A}_{j}}\,  a^{\prime \prime}\,  N{}^{(-5)} a{}^{(-3)} -\nonumber \\
	&& 3\, {A}_{0}  {\delta}^{i j} a^{\prime}\,  {\partial}_{i}{{A}_{j}}\,  N{}^{(-6)} N^\prime{}^{2} a{}^{(-3)} - 12\, {A}_{0}  {\partial}_{0}{{A}_{0}}\,  N^{\prime}\,  N{}^{(-7)} a^\prime{}^{2} a{}^{(-2)} + 6\, {A}_{0}  {\partial}_{0}{{A}_{0}}\,  N{}^{(-6)} a^\prime{}^{3} a{}^{(-3)} + \nonumber \\
	&&12\, {A}_{0}  {\partial}_{0}{{A}_{0}}\,  a^{\prime}\,  a^{\prime \prime}\,  N{}^{(-6)} a{}^{(-2)}%
	+ {A}_{0}  {\delta}^{i j} N^{\prime}\,  {\partial}_{i}{{A}_{j}}\,  N{}^{(-5)} a^\prime{}^{2} a{}^{(-4)} - 3\, {A}_{0}  {\delta}^{i j} a^{\prime}\,  {\partial}_{i}{{A}_{j}}\,  a^{\prime \prime}\,  N{}^{(-4)} a{}^{(-4)} - \nonumber \\
	&&{A}_{0}  {\delta}^{i j} a^{\prime}\,  {\partial}_{j}{{A}_{i}}\,  N{}^{(-6)} N^\prime{}^{2} a{}^{(-3)} + 7\, {A}_{0}  {\delta}^{i j} N^{\prime}\,  {\partial}_{j}{{A}_{i}}\,  N{}^{(-5)} a^\prime{}^{2} a{}^{(-4)} + {A}_{0}  {\delta}^{i j} N^{\prime}\,  {\partial}_{j}{{A}_{i}}\,  a^{\prime \prime}\,  N{}^{(-5)} a{}^{(-3)} + \nonumber \\
	&&3\,  a^{\prime \prime}\,  {A}_{0}{}^{2} N{}^{(-8)} N^\prime{}^{2} a{}^{-1} - 3\,  a^{\prime}\,  {A}_{0}{}^{2} N{}^{(-9)} N^\prime{}^{3} a{}^{-1} + 12\,  {A}_{0}{}^{2} N{}^{(-8)} N^\prime{}^{2} a^\prime{}^{2} a{}^{(-2)} - \nonumber \\
	&&15\,  N^{\prime}\,  {A}_{0}{}^{2} N{}^{(-7)} a^\prime{}^{3} a{}^{(-3)} - 12\,  N^{\prime}\,  a^{\prime}\,  a^{\prime \prime}\,  {A}_{0}{}^{2} N{}^{(-7)} a{}^{(-2)} - 12\, {A}_{0}  {\delta}^{i j} {\partial}_{j}{{A}_{i}}\,  N{}^{(-4)} a^\prime{}^{3} a{}^{(-5)} - \nonumber \\
	&&3\, {A}_{0}  {\delta}^{i j} a^{\prime}\,  {\partial}_{j}{{A}_{i}}\,  a^{\prime \prime}\,  N{}^{(-4)} a{}^{(-4)} + 18\,  {A}_{0}{}^{2} N{}^{(-6)} a^\prime{}^{4} a{}^{(-4)} + 9\,  a^{\prime \prime}\,  {A}_{0}{}^{2} N{}^{(-6)} a^\prime{}^{2} a{}^{(-3)}
	\end{eqnarray}
	\begin{eqnarray}
	&& \mathcal{L}_6 = 3\,  a^{\prime \prime}\,  N{}^{(-6)} {\partial}_{0}{{A}_{0}}\, {}^{2} a{}^{-1} - 3\,  N^{\prime}\,  a^{\prime}\,  N{}^{(-7)} {\partial}_{0}{{A}_{0}}\, {}^{2} a{}^{-1} - 3\,  {\delta}^{i j} {\partial}_{0}{{A}_{j}}\,  {\partial}_{i}{{A}_{0}}\,  a^{\prime \prime}\,  N{}^{(-4)} a{}^{(-3)} + \nonumber \\
	&&3\,  {\delta}^{i j} {\partial}_{0}{{A}_{j}}\,  N^{\prime}\,  a^{\prime}\,  {\partial}_{i}{{A}_{0}}\,  N{}^{(-5)} a{}^{(-3)} +  {\delta}^{i j} {\partial}_{0}{{A}_{i}}\,  N^{\prime}\,  a^{\prime}\,  {\partial}_{j}{{A}_{0}}\,  N{}^{(-5)} a{}^{(-3)} - 4\,  {\delta}^{i j} {\partial}_{0}{{A}_{i}}\,  {\partial}_{j}{{A}_{0}}\,  N{}^{(-4)} a^\prime{}^{2} a{}^{(-4)} - \nonumber \\
	&& {\delta}^{i j} {\partial}_{0}{{A}_{i}}\,  {\partial}_{j}{{A}_{0}}\,  a^{\prime \prime}\,  N{}^{(-4)} a{}^{(-3)} + 2\,  {\delta}^{i j} {\partial}_{0}{{A}_{j}}\,  {\partial}_{i}{{A}_{0}}\,  N{}^{(-4)} a^\prime{}^{2} a{}^{(-4)} -  {\delta}^{i j} {\delta}^{k l} N^{\prime}\,  a^{\prime}\,  {\partial}_{j}{{A}_{l}}\,  {\partial}_{k}{{A}_{i}}\,  N{}^{(-3)} a{}^{(-5)} + \nonumber \\
	&&4\,  {\delta}^{i j} {\delta}^{k l} {\partial}_{j}{{A}_{l}}\,  {\partial}_{k}{{A}_{i}}\,  N{}^{(-2)} a^\prime{}^{2} a{}^{(-6)} +  {\delta}^{i j} {\delta}^{k l} {\partial}_{j}{{A}_{l}}\,  {\partial}_{k}{{A}_{i}}\,  a^{\prime \prime}\,  N{}^{(-2)} a{}^{(-5)} - 2\,  {\delta}^{i j} {\delta}^{k l} {\partial}_{i}{{A}_{l}}\,  {\partial}_{k}{{A}_{j}}\,  N{}^{(-2)} a^\prime{}^{2} a{}^{(-6)} - \nonumber \\
	&&6\, {A}_{0}  {\partial}_{0}{{A}_{0}}\,  N^{\prime}\,  a^{\prime \prime}\,  N{}^{(-7)} a{}^{-1} + 6\, {A}_{0}  {\partial}_{0}{{A}_{0}}\,  a^{\prime}\,  N{}^{(-8)} N^\prime{}^{2} a{}^{-1} - {A}_{j}  {\delta}^{j i} N^{\prime}\,  {\partial}_{i}{{A}_{0}}\,  N{}^{(-5)} a^\prime{}^{2} a{}^{(-4)} + \nonumber \\
	&&2\, {A}_{j}  {\delta}^{j i} {\partial}_{i}{{A}_{0}}\,  N{}^{(-4)} a^\prime{}^{3} a{}^{(-5)} + {A}_{j}  {\delta}^{j i} a^{\prime}\,  {\partial}_{i}{{A}_{0}}\,  a^{\prime \prime}\,  N{}^{(-4)} a{}^{(-4)} + 3\, {A}_{j}  {\delta}^{j i} {\partial}_{0}{{A}_{i}}\,  a^{\prime}\,  a^{\prime \prime}\,  N{}^{(-4)} a{}^{(-4)} - \nonumber \\
	&&3\, {A}_{j}  {\delta}^{j i} {\partial}_{0}{{A}_{i}}\,  N^{\prime}\,  N{}^{(-5)} a^\prime{}^{2} a{}^{(-4)}%
	+ {A}_{0}  {\delta}^{i j} N^{\prime}\,  {\partial}_{i}{{A}_{j}}\,  N{}^{(-5)} a^\prime{}^{2} a{}^{(-4)} - {A}_{0}  {\delta}^{i j} a^{\prime}\,  {\partial}_{i}{{A}_{j}}\,  a^{\prime \prime}\,  N{}^{(-4)} a{}^{(-4)} + \nonumber \\
	&&3\, {A}_{i}  {\delta}^{i j} a^{\prime}\,  {\partial}_{j}{{A}_{0}}\,  a^{\prime \prime}\,  N{}^{(-4)} a{}^{(-4)} - 3\, {A}_{i}  {\delta}^{i j} N^{\prime}\,  {\partial}_{j}{{A}_{0}}\,  N{}^{(-5)} a^\prime{}^{2} a{}^{(-4)} + 3\,  a^{\prime \prime}\,  {A}_{0}{}^{2} N{}^{(-8)} N^\prime{}^{2} a{}^{-1} - \nonumber \\
	&&3\,  a^{\prime}\,  {A}_{0}{}^{2} N{}^{(-9)} N^\prime{}^{3} a{}^{-1} - 4\, {A}_{i} {A}_{j}  {\delta}^{i j} a^{\prime \prime}\,  N{}^{(-4)} a^\prime{}^{2} a{}^{(-5)} + 4\, {A}_{i} {A}_{j}  {\delta}^{i j} N^{\prime}\,  N{}^{(-5)} a^\prime{}^{3} a{}^{(-5)} - \nonumber \\
	&&{A}_{i}  {\delta}^{i j} {\partial}_{0}{{A}_{j}}\,  N^{\prime}\,  N{}^{(-5)} a^\prime{}^{2} a{}^{(-4)} + 2\, {A}_{i}  {\delta}^{i j} {\partial}_{0}{{A}_{j}}\,  N{}^{(-4)} a^\prime{}^{3} a{}^{(-5)} + {A}_{i}  {\delta}^{i j} {\partial}_{0}{{A}_{j}}\,  a^{\prime}\,  a^{\prime \prime}\,  N{}^{(-4)} a{}^{(-4)} - \nonumber \\
	&&4\, {A}_{i} {A}_{j}  {\delta}^{i j} N{}^{(-4)} a^\prime{}^{4} a{}^{(-6)} + 2\, {A}_{i} {A}_{j}  {\delta}^{j i} N{}^{(-4)} a^\prime{}^{4} a{}^{(-6)} + {A}_{0}  {\delta}^{i j} N^{\prime}\,  {\partial}_{j}{{A}_{i}}\,  N{}^{(-5)} a^\prime{}^{2} a{}^{(-4)} - \nonumber \\
	&&4\, {A}_{0}  {\delta}^{i j} {\partial}_{j}{{A}_{i}}\,  N{}^{(-4)} a^\prime{}^{3} a{}^{(-5)} - {A}_{0}  {\delta}^{i j} a^{\prime}\,  {\partial}_{j}{{A}_{i}}\,  a^{\prime \prime}\,  N{}^{(-4)} a{}^{(-4)} - 3\,  N^{\prime}\,  {A}_{0}{}^{2} N{}^{(-7)} a^\prime{}^{3} a{}^{(-3)} + \nonumber \\
	&&6\,  {A}_{0}{}^{2} N{}^{(-6)} a^\prime{}^{4} a{}^{(-4)} + 3\,  a^{\prime \prime}\,  {A}_{0}{}^{2} N{}^{(-6)} a^\prime{}^{2} a{}^{(-3)}
	\end{eqnarray}
	\begin{eqnarray}
	&& \mathcal{L}_7 = - {\delta}^{i j} {\partial}_{0}{{A}_{i}}\,  N^{\prime}\,  a^{\prime}\,  {\partial}_{j}{{A}_{0}}\,  N{}^{(-5)} a{}^{(-3)} + {\delta}^{i j} {\partial}_{0}{{A}_{i}}\,  {\partial}_{j}{{A}_{0}}\,  a^{\prime \prime}\,  N{}^{(-4)} a{}^{(-3)} + {\delta}^{i j} {\partial}_{0}{{A}_{i}}\,  {\partial}_{0}{{A}_{j}}\,  N^{\prime}\,  a^{\prime}\,  N{}^{(-5)} a{}^{(-3)} - \nonumber \\
	&&{\delta}^{i j} {\partial}_{0}{{A}_{i}}\,  {\partial}_{0}{{A}_{j}}\,  a^{\prime \prime}\,  N{}^{(-4)} a{}^{(-3)} + {\delta}^{i j} N^{\prime}\,  a^{\prime}\,  {\partial}_{i}{{A}_{0}}\,  {\partial}_{j}{{A}_{0}}\,  N{}^{(-5)} a{}^{(-3)} - {\delta}^{i j} {\partial}_{i}{{A}_{0}}\,  {\partial}_{j}{{A}_{0}}\,  a^{\prime \prime}\,  N{}^{(-4)} a{}^{(-3)} - \nonumber \\
	&&{\delta}^{i j} {\partial}_{0}{{A}_{j}}\,  N^{\prime}\,  a^{\prime}\,  {\partial}_{i}{{A}_{0}}\,  N{}^{(-5)} a{}^{(-3)} + {\delta}^{i j} {\partial}_{0}{{A}_{j}}\,  {\partial}_{i}{{A}_{0}}\,  a^{\prime \prime}\,  N{}^{(-4)} a{}^{(-3)} + {\delta}^{i j} {\delta}^{k l} {\partial}_{k}{{A}_{i}}\,  {\partial}_{l}{{A}_{j}}\,  N{}^{(-2)} a^\prime{}^{2} a{}^{(-6)} - \nonumber \\
	&&{\delta}^{i j} {\delta}^{k l} {\partial}_{j}{{A}_{l}}\,  {\partial}_{k}{{A}_{i}}\,  N{}^{(-2)} a^\prime{}^{2} a{}^{(-6)}
	\end{eqnarray}}

{\small
	\begin{eqnarray}
	&&\mathcal{L}_8 = - {\delta}^{i j} {\partial}_{0}{{A}_{0}}\,  N^{\prime}\,  a^{\prime}\,  {\partial}_{j}{{A}_{i}}\,  N{}^{(-5)} a{}^{(-3)} + {\delta}^{i j} {\partial}_{0}{{A}_{0}}\,  {\partial}_{j}{{A}_{i}}\,  a^{\prime \prime}\,  N{}^{(-4)} a{}^{(-3)} + {\delta}^{i j} {\partial}_{0}{{A}_{i}}\,  {\partial}_{0}{{A}_{j}}\,  N^{\prime}\,  a^{\prime}\,  N{}^{(-5)} a{}^{(-3)} - \nonumber \\
	&& {\delta}^{i j} {\partial}_{0}{{A}_{i}}\,  {\partial}_{0}{{A}_{j}}\,  a^{\prime \prime}\,  N{}^{(-4)} a{}^{(-3)} + {\delta}^{i j} N^{\prime}\,  a^{\prime}\,  {\partial}_{i}{{A}_{0}}\,  {\partial}_{j}{{A}_{0}}\,  N{}^{(-5)} a{}^{(-3)} - {\delta}^{i j} {\partial}_{i}{{A}_{0}}\,  {\partial}_{j}{{A}_{0}}\,  a^{\prime \prime}\,  N{}^{(-4)} a{}^{(-3)} -\nonumber \\
	&& {\delta}^{i j} {\partial}_{0}{{A}_{0}}\,  N^{\prime}\,  a^{\prime}\,  {\partial}_{i}{{A}_{j}}\,  N{}^{(-5)} a{}^{(-3)} + {\delta}^{i j} {\partial}_{0}{{A}_{0}}\,  {\partial}_{i}{{A}_{j}}\,  a^{\prime \prime}\,  N{}^{(-4)} a{}^{(-3)} + {\delta}^{i j} {\delta}^{k l} {\partial}_{k}{{A}_{j}}\,  {\partial}_{l}{{A}_{i}}\,  N{}^{(-2)} a^\prime{}^{2} a{}^{(-6)} -\nonumber \\
	&& {\delta}^{i j} {\delta}^{k l} {\partial}_{j}{{A}_{i}}\,  {\partial}_{k}{{A}_{l}}\,  N{}^{(-2)} a^\prime{}^{2} a{}^{(-6)} - 3\, {A}_{i} {\delta}^{i j} {\partial}_{0}{{A}_{j}}\,  N^{\prime}\,  N{}^{(-5)} a^\prime{}^{2} a{}^{(-4)} + 2\, {A}_{i} {\delta}^{i j} {\partial}_{0}{{A}_{j}}\,  a^{\prime}\,  a^{\prime \prime}\,  N{}^{(-4)} a{}^{(-4)} + \nonumber \\
	&&{A}_{j} {\delta}^{j i} {\partial}_{0}{{A}_{i}}\,  N^{\prime}\,  N{}^{(-5)} a^\prime{}^{2} a{}^{(-4)} + {A}_{0} {\delta}^{i j} a^{\prime}\,  {\partial}_{i}{{A}_{j}}\,  N{}^{(-6)} N^\prime{}^{2} a{}^{(-3)} - {A}_{0} {\delta}^{i j} N^{\prime}\,  {\partial}_{i}{{A}_{j}}\,  a^{\prime \prime}\,  N{}^{(-5)} a{}^{(-3)} - \nonumber \\
	&&{A}_{j} {\delta}^{j i} N^{\prime}\,  {\partial}_{i}{{A}_{0}}\,  N{}^{(-5)} a^\prime{}^{2} a{}^{(-4)} + {A}_{j} {\delta}^{j i} a^{\prime}\,  {\partial}_{i}{{A}_{0}}\,  a^{\prime \prime}\,  N{}^{(-4)} a{}^{(-4)} + 6\, {A}_{0} {\partial}_{0}{{A}_{0}}\,  N^{\prime}\,  N{}^{(-7)} a^\prime{}^{2} a{}^{(-2)} - \nonumber \\
	&&6\, {A}_{0} {\partial}_{0}{{A}_{0}}\,  a^{\prime}\,  a^{\prime \prime}\,  N{}^{(-6)} a{}^{(-2)}%
	+ 2\, {A}_{0} {\delta}^{i j} {\partial}_{i}{{A}_{j}}\,  N{}^{(-4)} a^\prime{}^{3} a{}^{(-5)} + {A}_{0} {\delta}^{i j} a^{\prime}\,  {\partial}_{j}{{A}_{i}}\,  N{}^{(-6)} N^\prime{}^{2} a{}^{(-3)} - \nonumber \\
	&&{A}_{0} {\delta}^{i j} N^{\prime}\,  {\partial}_{j}{{A}_{i}}\,  a^{\prime \prime}\,  N{}^{(-5)} a{}^{(-3)} + {A}_{i} {A}_{j} {\delta}^{i j} N^{\prime}\,  N{}^{(-5)} a^\prime{}^{3} a{}^{(-5)} - {A}_{i} {A}_{j} {\delta}^{i j} a^{\prime \prime}\,  N{}^{(-4)} a^\prime{}^{2} a{}^{(-5)} - \nonumber \\
	&&6\, {A}_{0}{}^{2} N{}^{(-8)} N^\prime{}^{2} a^\prime{}^{2} a{}^{(-2)} + 6\, N^{\prime}\,  a^{\prime}\,  a^{\prime \prime}\,  {A}_{0}{}^{2} N{}^{(-7)} a{}^{(-2)} - {A}_{i} {\delta}^{i j} N^{\prime}\,  {\partial}_{j}{{A}_{0}}\,  N{}^{(-5)} a^\prime{}^{2} a{}^{(-4)} + \nonumber \\
	&&{A}_{i} {\delta}^{i j} a^{\prime}\,  {\partial}_{j}{{A}_{0}}\,  a^{\prime \prime}\,  N{}^{(-4)} a{}^{(-4)} + {A}_{i} {A}_{j} {\delta}^{j i} N^{\prime}\,  N{}^{(-5)} a^\prime{}^{3} a{}^{(-5)} - {A}_{i} {A}_{j} {\delta}^{j i} a^{\prime \prime}\,  N{}^{(-4)} a^\prime{}^{2} a{}^{(-5)} + \nonumber \\
	&&2\, {A}_{0} {\delta}^{i j} {\partial}_{j}{{A}_{i}}\,  N{}^{(-4)} a^\prime{}^{3} a{}^{(-5)} - 6 {A}_{0}{}^{2} N{}^{(-6)} a^\prime{}^{4} a{}^{(-4)}
	\end{eqnarray}
	\begin{eqnarray}
	&&\mathcal{L}_9 = - 36\, {A}_{0}{}^{2} N{}^{(-6)} a^{\prime \prime}{}^{2} a{}^{(-2)} + 72\, N^{\prime}\,  a^{\prime}\,  a^{\prime \prime}\,  {A}_{0}{}^{2} N{}^{(-7)} a{}^{(-2)} - 36\, {A}_{0}{}^{2} N{}^{(-8)} N^\prime{}^{2} a^\prime{}^{2} a{}^{(-2)} + \nonumber \\
	&&36\, {A}_{i} {A}_{j} {\delta}^{i j} N{}^{(-4)} a^{\prime \prime}{}^{2} a{}^{(-4)} - 72\, {A}_{i} {A}_{j} {\delta}^{i j} N^{\prime}\,  a^{\prime}\,  a^{\prime \prime}\,  N{}^{(-5)} a{}^{(-4)} + 36\, {A}_{i} {A}_{j} {\delta}^{i j} N{}^{(-6)} N^\prime{}^{2} a^\prime{}^{2} a{}^{(-4)} - \nonumber \\
	&&72\, a^{\prime \prime}\,  {A}_{0}{}^{2} N{}^{(-6)} a^\prime{}^{2} a{}^{(-3)} + 72\, N^{\prime}\,  {A}_{0}{}^{2} N{}^{(-7)} a^\prime{}^{3} a{}^{(-3)} + 72\, {A}_{i} {A}_{j} {\delta}^{i j} a^{\prime \prime}\,  N{}^{(-4)} a^\prime{}^{2} a{}^{(-5)} - \nonumber \\
	&&72\, {A}_{i} {A}_{j} {\delta}^{i j} N^{\prime}\,  N{}^{(-5)} a^\prime{}^{3} a{}^{(-5)} - 36\, {A}_{0}{}^{2} N{}^{(-6)} a^\prime{}^{4} a{}^{(-4)} + 36\, {A}_{i} {A}_{j} {\delta}^{i j} N{}^{(-4)} a^\prime{}^{4} a{}^{(-6)}
	\end{eqnarray}
	\begin{eqnarray}
	&&\mathcal{L}_{10} = - 18\, {A}_{0}{}^{2} N{}^{(-6)} a^{\prime \prime}{}^{2} a{}^{(-2)} + 36\, N^{\prime}\,  a^{\prime}\,  a^{\prime \prime}\,  {A}_{0}{}^{2} N{}^{(-7)} a{}^{(-2)} - 18\, {A}_{0}{}^{2} N{}^{(-8)} N^\prime{}^{2} a^\prime{}^{2} a{}^{(-2)} - \nonumber \\
	&&12\, {A}_{i} {A}_{j} {\delta}^{i j} N^{\prime}\,  a^{\prime}\,  a^{\prime \prime}\,  N{}^{(-5)} a{}^{(-4)} + 18\, {A}_{i} {A}_{j} {\delta}^{i j} a^{\prime \prime}\,  N{}^{(-4)} a^\prime{}^{2} a{}^{(-5)} + 6\, {A}_{i} {A}_{j} {\delta}^{i j} N{}^{(-4)} a^{\prime \prime}{}^{2} a{}^{(-4)} + \nonumber \\
	&&6\, {A}_{i} {A}_{j} {\delta}^{i j} N{}^{(-6)} N^\prime{}^{2} a^\prime{}^{2} a{}^{(-4)} - 18\, {A}_{i} {A}_{j} {\delta}^{i j} N^{\prime}\,  N{}^{(-5)} a^\prime{}^{3} a{}^{(-5)} - 18\, a^{\prime \prime}\,  {A}_{0}{}^{2} N{}^{(-6)} a^\prime{}^{2} a{}^{(-3)} + \nonumber \\
	&&18\, N^{\prime}\,  {A}_{0}{}^{2} N{}^{(-7)} a^\prime{}^{3} a{}^{(-3)} + 12\, {A}_{i} {A}_{j} {\delta}^{i j} N{}^{(-4)} a^\prime{}^{4} a{}^{(-6)} 
	\end{eqnarray}
	\begin{eqnarray}
	&&\mathcal{L}_{11} = - 12\, {A}_{0}{}^{2} N{}^{(-6)} a^{\prime \prime}{}^{2} a{}^{(-2)} + 24\, N^{\prime}\,  a^{\prime}\,  a^{\prime \prime}\,  {A}_{0}{}^{2} N{}^{(-7)} a{}^{(-2)} - 12\, {A}_{0}{}^{2} N{}^{(-8)} N^\prime{}^{2} a^\prime{}^{2} a{}^{(-2)} + \nonumber \\
	&&12\, {A}_{i} {A}_{j} {\delta}^{i j} N{}^{(-4)} a^{\prime \prime}{}^{2} a{}^{(-4)} - 24\, {A}_{i} {A}_{j} {\delta}^{i j} N^{\prime}\,  a^{\prime}\,  a^{\prime \prime}\,  N{}^{(-5)} a{}^{(-4)} + 12\, {A}_{i} {A}_{j} {\delta}^{i j} N{}^{(-6)} N^\prime{}^{2} a^\prime{}^{2} a{}^{(-4)} + \nonumber \\
	&&12\, N^{\prime}\,  {A}_{0}{}^{2} N{}^{(-7)} a^\prime{}^{3} a{}^{(-3)} - 12\, {A}_{0}{}^{2} N{}^{(-6)} a^\prime{}^{4} a{}^{(-4)} - 12\, a^{\prime \prime}\,  {A}_{0}{}^{2} N{}^{(-6)} a^\prime{}^{2} a{}^{(-3)} - \nonumber \\
	&&12\, {A}_{i} {A}_{j} {\delta}^{i j} N^{\prime}\,  N{}^{(-5)} a^\prime{}^{3} a{}^{(-5)} + 12\, {A}_{i} {A}_{j} {\delta}^{i j} N{}^{(-4)} a^\prime{}^{4} a{}^{(-6)} + 12\, {A}_{i} {A}_{j} {\delta}^{i j} a^{\prime \prime}\,  N{}^{(-4)} a^\prime{}^{2} a{}^{(-5)} 
	\end{eqnarray}
	\begin{eqnarray}
	&&\mathcal{L}_{12} = - 9\, {A}_{0}{}^{2} N{}^{(-6)} a^{\prime \prime}{}^{2} a{}^{(-2)} + 18\, N^{\prime}\,  a^{\prime}\,  a^{\prime \prime}\,  {A}_{0}{}^{2} N{}^{(-7)} a{}^{(-2)} - 9\, {A}_{0}{}^{2} N{}^{(-8)} N^\prime{}^{2} a^\prime{}^{2} a{}^{(-2)} + \nonumber \\
	&&{A}_{i} {A}_{j} {\delta}^{i j} N{}^{(-6)} N^\prime{}^{2} a^\prime{}^{2} a{}^{(-4)} - 4\, {A}_{i} {A}_{j} {\delta}^{i j} N^{\prime}\,  N{}^{(-5)} a^\prime{}^{3} a{}^{(-5)} - 2\, {A}_{i} {A}_{j} {\delta}^{i j} N^{\prime}\,  a^{\prime}\,  a^{\prime \prime}\,  N{}^{(-5)} a{}^{(-4)} + \nonumber \\
	&&4\, {A}_{i} {A}_{j} {\delta}^{i j} N{}^{(-4)} a^\prime{}^{4} a{}^{(-6)} + 4\, {A}_{i} {A}_{j} {\delta}^{i j} a^{\prime \prime}\,  N{}^{(-4)} a^\prime{}^{2} a{}^{(-5)} + {A}_{i} {A}_{j} {\delta}^{i j} N{}^{(-4)} a^{\prime \prime}{}^{2} a{}^{(-4)}
	\end{eqnarray}
}

where the action is defined as:
\begin{eqnarray}
\mathcal{S}_i &=&  D_i\int d^4 x \sqrt{-g}\,\mathcal{L}_i \nonumber \\
&=& D_i\int d^4x\,N\,a^3\,\mathcal{L}_i
\end{eqnarray}

\section{Spectrum of electric and $B.B^\prime$}\label{app:otherEner}

Electric part and $B.B^\prime$ part of the energy density at sound horizon become

\begin{eqnarray}
\mathcal{P}_E &=& -
\frac{24\,D}{\pi\,c_s^{7+2\beta}}\,\mathcal{G}_1(\beta)\,H_*^4\,\left(\frac{k}{k_*}\right)^{2\beta
  + 8}, \beta \langle -\frac{5}{2},~~\mathcal{G}_1(\beta) =
\frac{|C_1|^2}{2^{2\beta+3}\left(\Gamma(\beta + 5/2)\right)^2}
\nonumber \\ && -
\frac{24\,D}{\pi\,c_s^{1-2\beta}}\,\mathcal{G}_2(\beta)\,H_*^4\,\left(\frac{k}{k_*}\right)^{2
  - 2\beta}, \beta \rangle -\frac{5}{2},~~\mathcal{G}_2(\beta) =
\frac{|C_2|^2}{2^{ -2\beta-3}\left(\Gamma( - \beta -1/2)\right)^2}
\\ \mathcal{P}_{B.B^\prime} &=& -\frac{16\,D}{\pi\,c_s^{10+2\beta}}
\mathcal{J}_1(\beta)\,H_*^4 \left(\frac{k}{k_*}\right)^{2\beta + 10},
\beta\langle -\frac{5}{2}, \mathcal{J}_1(\beta) =
\frac{|C_1|^2}{2^{2\beta+4} (-\beta - 5/2)\,\left(\Gamma(\beta +
  5/2)\right)^2} \nonumber \\ && - \frac{16\,D}{\pi\,c_s^{2 - 2
    \beta}} \mathcal{J}_2(\beta)\,H_*^4 \left(\frac{ k}{k_*}\right)^{2
  -2\beta}, \beta \rangle -\frac{5}{2}, \mathcal{J}_2(\beta) =
\frac{|C_2|^2}{2^{-2\beta - 4} (-\beta - 3/2)\,\left(\Gamma(-\beta -
  3/2)\right)^2}
\end{eqnarray}

\section{Slow-roll inflation and spectrum of the energy densities}
\label{app:slow-roll}

In case of slow-roll inflation, the slow-roll parameters are defined
as
\begin{eqnarray}
&&\epsilon_1 = -\frac{\dot{H}}{H^2},~~ \mbox{where dot used for cosmic time }t \nonumber \\
&&\epsilon_{n+1} = \frac{\dot{\epsilon}_n}{H\epsilon}, ~~n~\mbox{is natural numbers.}
\end{eqnarray}

In conformal coordinate, 
\begin{eqnarray}
&&\epsilon_1 - 1 = -\frac{\mathcal{H}^\prime}{\mathcal{H}^2} \\
\Rightarrow&& c_s \equiv \frac{\mathcal{H}^\prime}{\mathcal{H}^2} = 1 - \epsilon_1 ~~~~\mbox{(exact)} \\
\end{eqnarray}
and
\begin{eqnarray}
\frac{J^{\prime\prime}}{J} = \mathcal{H}^2\left(-\epsilon_1 + 2 \epsilon_1^2 - \epsilon_1\epsilon_2\right)~~~~\mbox{(exact)}
\end{eqnarray} 

In case of leading order slow-roll approximation with $\epsilon_n \ll 1$,  $\mathcal{H} = -\frac{1+\epsilon_1}{\eta}$, thus
\begin{eqnarray}
\frac{J^{\prime\prime}}{J} = \frac{\epsilon_1(\epsilon_2 - 1)}{\eta^2}
\end{eqnarray}

Hence, the equation for $\mathcal{A}_k$ becomes
\begin{eqnarray}
\mathcal{A}_k^{\prime\prime} + \left(c_s^2 k^2 - \frac{\epsilon_1(\epsilon_2 - 1)}{\eta^2}\right)\,\mathcal{A}_k = 0
\end{eqnarray}

Solution for the above equation using Bunch-Davies vacuum becomes
\begin{eqnarray}
\mathcal{A}_k = \frac{\pi}{4} \sqrt{-\eta} H^1_\nu (-c_s k \eta), ~~\nu = \frac{1}{2}\sqrt{1 + \epsilon_1(\epsilon_2 - 1)}
\end{eqnarray}

Using the above solution for $-k\eta \rightarrow 0$, at sound horizon, spectral energy density of the magnetic field becomes
\begin{eqnarray}
\mathcal{P}_B = \frac{8\,D}{\pi\,c_s^5} \,H_*^4 \left(\frac{k}{k_*}\right)^4
\end{eqnarray}

\providecommand{\href}[2]{#2}\begingroup\raggedright\endgroup

\end{document}